\documentclass[aps,prd,twocolumn,%notitlepage,
nofootinbib]{revtex4-2}

\usepackage[utf8]{inputenc}
\inputencoding{utf8}

\usepackage{amsmath, amssymb,mathrsfs}
\usepackage{epsfig,color,graphicx}
\usepackage{xcolor}
\usepackage{subfigure}
\colorlet{darkred}{red!80!black}
\colorlet{darkgreen}{green!50!black}
\colorlet{darkblue}{blue!50!black}

\usepackage{soul}

\allowdisplaybreaks

\newcommand{\cd}{\makebox[0.08cm]{$\cdot$}}

\begin{document}

\title{Restoration of rotational symmetry in light-front dynamics}

\author{V.A.~Karmanov} 
\email{karmanovva@lebedev.ru}
\affiliation{Lebedev Physical Institute, Leninsky Prospekt 53, 119991 Moscow, Russia}

\author{E.A. Kupriyanova}
\email{kupriyanovaea@lebedev.ru, kupr\_i\_k@mail.ru}
\affiliation{Lebedev Physical Institute, Leninsky Prospekt 53, 119991 Moscow, Russia}

\date{\today}

\begin{abstract}
Solving the bound state problem in the framework of the light-front dynamics (with the state vector defined in the light-front plane $t+z=0$), one finds that after truncation - representing approximately the state vector by a finite number of the Fock components, the binding energies become to depend on projection of the angular momentum on the $z$-axes. This is so called violation of rotational symmetry in light-front dynamics. It is shown that the physical solution is the superposition of these states with some well defined coefficients. For a test, we also
solve the Bethe-Salpeter equation and projecting its solution on the light-front, we find “true” 
wave functions which don’t suffer from violation of the rotation invariance. Comparison with them of solution in the form of superposition shows that not only binding energies but also the wave functions themselves are described by the superpositions of the split states fairly well.
\end{abstract}
\maketitle

\section{Introduction}\label{intro}
Light-front dynamics (LFD), proposed in \cite{Dirac}, is now a powerful and efficient approach to study the relativistic few-body bound systems (nuclei and hadrons), in particular, in the QCD framework \cite{WP}. It has many advantages. One of them is in the fact that it deals with the 
physically transparent objects - the wave functions (WF), having the probability interpretation. See for review and applications 
Refs. \cite{Coester92,cdkm,Brodsky98}. 

The relativistic light-front wave function (LFWF) is defined in the three-dimensional subspace  
 of the 4D Minkowski space - on the LF plane  $t+z=0$  (we put the speed of light $c=1$). Though in this equation the axe $z$ is distinguished from the $x,y$ axes, 
this does not lead to violation of the rotational symmetry,  if algebra of the Poincar\'e operators (commutation relations) is preserved. Similarly, the spherical functions $Y_{lm}(\hat{\vec{r}})$ with definite projections $m$ of angular momentum on the $z$-axis (distinguished from the $x,y$ axes) perfectly agrees with the rotational symmetry. 
However, in practice, when we make approximations (for example, we 
truncate the Fock space, omitting the higher sectors), this is equivalent to implicit modifying the Poincar\'e group generators $J_{ij}$, defined relative to the axes $i,j$. Some axes become really distinguished, the equivalence of the $z$- and $x,y$-axes is violated. As a result, the mass of the system becomes to depend on projection $m$ of the angular momentum on the $z$-axe. 
 This is just a manifestation of the notorious violation of the rotational symmetry in LFD. 
 Solving the eigenvalue problem, we find whole spectrum of the  masses depending on the projection $m$.
 Which mass of them and associated with it eigenfunction -- LFWF  should we put in correspondence to the physical system which we describe?

We emphasize again that this violation is not a physical phenomenon. It is an artefact of truncation. 
We imply that  behind this picture there is a quantum field theory. When the generators are phenomenologically constructed not from the field operators but from the particle coordinates or momenta \cite{Coester92,Polyzou}, they can be built so that the Poincar\'e algebra is satisfied even in the two-body (or three-body) sector that saves the rotational symmetry.

This problem remains, but it manifests itself in a different form in the framework of
 the explicitly covariant version of LFD (see for review  \cite{cdkm}). In this version, the state vector is defined on the LF plane of general orientation, determined by the equation $\omega\cd x=0$, where $\omega$ is a four-vector 
$\omega=(\omega_0,\vec{\omega})$ such that $\omega^2=0$.  Standard LFD is recovered as a particular case, for $\omega=(1,0,0,-1)$. 

In the explicitly covariant LFD violation of the rotational symmetry obtains a little bit different form.
Transformations of the state vector under rotation of the reference frame and under rotation of the LF plane $\omega\cd x=0$ around the light cone in a given reference frame are independent from each other. 
As usual, the states are labeled by the total total angular momentum $J$ and its projection $m$ on the $z$-axes. However, simultaneously, as we will see below, before taking into account some special constraint, sometimes called the angular condition, the states are labeled by projection $a$ of the angular momentum on the direction $\vec{\omega}$. That is, by the eigenvalues of the operator $A^2=(\vec{\omega}\vec{J})^2$. In this operator the two vectors 
$\vec{\omega}$ and $\vec{J}$ are transformed  under rotations in the same way. $A$ is scalar and therefore it commutes with $\vec{J}^2,\,J_z$. The operator $A$ itself has the integer eigenvalues $-l \leq a\leq l$. We take $A^2$ for convenience,  in order to deal with the positive eigenvalues $0 \leq a\leq l$.
We get a set of solutions for LFWF, labeled, in addition to $lm$, by $a$. Then the same question arises:  LFWF with which $a$  should we put in correspondence to the physical system which we describe?  In the exact case, these states are degenerate. The corresponding masses $M_a$ are the same, 
but the wave functions are different. Then the same question is related not to the masses $M_a$ only, but to the wave functions as well.  After truncation the
masses $M_a$ don’t depend on the projection $m$, but depend on $a$. 

Our answer is: we should put in correspondence with the physical system a superposition of the states with the different $a$.  When, after truncation,
the states in the superposition ceases to be degenerate, then this superposition becomes to be a little bit unusual -  is a sum of states with the different energies. It is an approximate eigenfunction both of the initial and of truncated Hamiltonian. 
The physical mass squared $M^2$ is the average value of the $a$-dependent masses $M_a^2$.
 Since the truncated Hamiltonian is  the approximate one, we are not motivated to find its exact eigenfunctions. However, we will compare the masses and the wave functions found in this way with ones found in completely different and independent way, namely, by projecting the Bethe-Salpeter (BS) amplitude on the light-front plane. 
 We will find that they are rather close to each other.
Below we will justify these statements and illustrate them by the numerical examples of the systems of two spinless particles with the total angular momentum $l=1$ and $l=2$.

\section{Equation for the two-body LFWF}\label{equation}
In this section we illustrate the problem by a simple example of two spinless particles interacting by exchange of the spinless particle. In the "ordinary" LFD, the
equation for the two-body LFWF (Weinberg equation \cite{Weinberg}) reads:
\begin{widetext}
\begin{equation}\label{eq1}
\left(\frac{\vec{R}^2_{\perp}+m^2}{x(1-x)}-M^2\right)
\psi(\vec{R}_{\perp},x)=
-\frac{m^2}{2\pi^3}\int\psi(\vec{R}'_{\perp},x')
V(\vec{R}'_{\perp},x';\vec{R}_{\perp},x,M^2)
\frac{d^2R'_{\perp}dx'}{2x'(1-x')}\ ,
\end{equation}
\end{widetext}
where $\vec{R}_{\perp},x$ are the standard LF transverse and longitudinal variables (we borrow the notation $\vec{R}_{\perp}$ from \cite{K-S}), $m$ is mass of  each constituent, $M=2m-B$ is the bound state mass, $B=|E_b|$ is the binding energy. The one-boson exchange kernel $V(\vec{R}'_{\perp},x';\vec{R}_{\perp},x,M^2)$, for $x \leq x'$, has the form (see e.g. Eq. (7.27) in \cite{cdkm}):
\begin{widetext}
\begin{equation}\label{kernrx}
V(\vec{R}'_{\perp},x';\vec{R}_{\perp},x,M^2)=-\frac{4\pi\alpha}
{\mu^2+\frac{x'}{x}\left(1-\frac{x}{x'}\right)^2m^2
+\frac{x'}{x}\left(\vec{R}_{\perp}-\frac{x}{x'}\vec{R}'_{\perp}\right)^2
+(x’-x)\left(\frac{m^2+\vec{R’}_{\perp}^2}{x’(1-x’)}-M^2\right)} \ ,
\end{equation}
\end{widetext}
$\mu$ is mass of the exchanged particle.
For $x \geq x'$ $V(\vec{R}'_{\perp},x';\vec{R}_{\perp},x,M^2)$ is obtained from
(\ref{kernrx}) by the replacement: $x \leftrightarrow x'$, $\vec{R}_{\perp}
\leftrightarrow \vec{R}'_{\perp}$.
The vector $\vec{R}_{\perp}$ is orthogonal to the $z$-axe, that is, it has the components $R_{x,y}$, whereas $R_z=0$. 
The equations (\ref{eq1}), (\ref{kernrx}) were initially derived in the standard LFD with the LF plane $t+z=0$ (more precisely, in the infinite momentum frame \cite{Weinberg}). They obviously do not obey three-dimensional rotational symmetry.

\begin{figure}[!ht]
\centering
\includegraphics[width=0.5\textwidth]{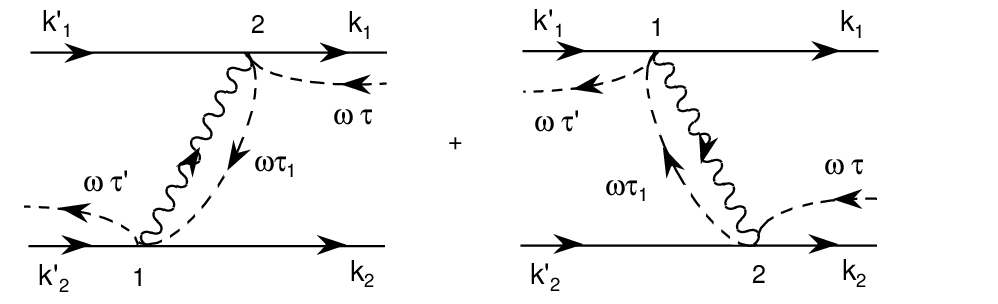}
\caption[*]{LF time ordered one boson exchange  graphs, shown in terms of the LF graph techniques presented in \cite{cdkm}.}
\label{obe_lf}
\end{figure}

In the explicitly covariant version of LFD \cite{cdkm}, when the wave function  is defined on the LF plane of general orientation $\omega\cd x=0$, where 
$\omega=(\omega_0,\vec{\omega})$ with $\omega^2=0$, the equations (\ref{eq1}) for LFWF turns into
\begin{widetext}
\begin{equation}\label{eqwf3a}
\left(4(\vec{k}\,^2 + m^2)-M^2\right)\psi(\vec{k},\vec{n}) =
-\frac{m^2}{2\pi^3} \int \psi(\vec{k}\,',\vec{n})
V(\vec{k}\,',\vec{k},\vec{n},M^2) \frac{d^3k'}{\varepsilon_{k'}} \ .
\end{equation}
One-boson exchange kernel (\ref{kernrx}) is graphically shown in Fig. \ref{obe_lf}.
It obtains the form:
 \begin{equation}\label{k3a}
V(\vec{k}\,',\vec{k},\vec{n},M^2)=-\frac{4\pi\alpha}
{\mu^2+ (\vec{k}\,' - \vec{k}\,)^2                                         -
(\vec{n}\cd \vec{k'}) (\vec{n}\cd \vec{k})\frac{(\varepsilon_{k'}-\varepsilon_k)^2}{\varepsilon_{k'}\varepsilon_k} 
+\left(\varepsilon_{k'}^2+\varepsilon_k^2-\frac{1}{2}M^2\right)
\left|\frac{(\vec{n}\cd \vec{k'})}{\varepsilon_{k'}}-\frac{(\vec{n}\cd \vec{k})}{\varepsilon_k}\right|}.
\end{equation}
\end{widetext}
Here $\vec{n}$  is the unit vector in the direction of $\vec{\omega}$.  In the standard LFD there is 
no conservation law for the “minus”-components of the momenta  defined as $k_-=k_0-k_z$. It is similar to the fact that in the instant form ($t=0$) there is 
no conservation law for the zero-components of the momenta  $k_0$ (for the energies): $k_{10}+k_{20}\neq p_0$.  In the explicitly covariant version, the momentum conservation obtains the form: $\vec{k}_1+\vec{k}_2=\vec{p}+\vec{\omega}\tau$.
For convenience, we work in the reference frame where $\vec{k}_1+\vec{k}_2=0\neq \vec{p}$,  $\vec{k}_1,\vec{k}_2$ are the constituent momenta and sum of them is not equal to the total momentum $\vec{p}$.

The equations (\ref{eq1}), (\ref{kernrx}) and (\ref{eqwf3a}), (\ref{k3a}) are equivalent to each other and they are transformed into each other  by the following replacement of variables:
\begin{equation}\label{Rx}
\vec{R}_{\perp}=\vec{k}-\vec{n}(\vec{n}\cd\vec{k}),\quad x=\frac{1}{2}\left(1-\frac{(\vec{n}\cd\vec{k})}{\varepsilon_k}\right),
\end{equation}
with $\varepsilon_k=\sqrt{m^2+\vec{k}^2}$. 

Equations in the form (\ref{eqwf3a}), (\ref{k3a}) are convenient for analyzing the non-relativistic limit.
For $k,k' \ll m$, the kernel (\ref{k3a}) turns into the Yukawa potential in the momentum space
\begin{equation}\label{k5}
V(\vec{k}\,',\vec{k})\simeq -\frac{4\pi\alpha}{\mu^2+(\vec{k}\,'-\vec{k})^2}
\end{equation}
corresponding to 
\begin{equation}\label{k5a}
V(r)=-\frac{\alpha}{r}\exp(-\mu r) 
\end{equation}
in the coordinate space.

The Schr\"odinger equation with the potential (\ref{k5}) or (\ref{k5a}) provides 
set of solutions
\begin{equation}\label{psi0}
\psi_{lm}(\vec{k})=g_l(k)Y_{lm}(\hat{\vec{k}}),
\end{equation}
($Y_{lm}(\hat{\vec{k}})$ is the spherical harmonics, $\hat{\vec{k}}$ is the unit vector in the direction $\vec{k}$)
 with the binding energies depending on the principal quantum number $n$ and on the angular momentum $l$. The binding energies don't depend on the projections $m$ of the angular momentum on the $z$-axe (they are degenerate relative to $m$). Corresponding angular momentum operator, whose eigenfunctions satisfy the Schr\"odinger equation with the kernel $V(\vec{k}\,',\vec{k})$  has the usual form
\begin{equation}\label{vecL}
\vec{L}_{\vec{k}}=-i\Bigl[\vec{k}\times \frac{\partial}{\partial \vec{k}}\Bigr]. 
\end{equation}
However, since the kernel (\ref{k3a}) depends on $\vec{n}$, the operator (\ref{vecL}) does not commute with  this kernel and therefore solutions of Eq. (\ref{eqwf3a}) are not anymore eigenstates of $\vec{L}^2_{\vec{k}}$. In other words, if $l(l+1)$ is eigenvalue of the operator 
$\vec{L}^2_{\vec{k}}$, then solutions of Eq. (\ref{eqwf3a}) don’t corresponds to definite $l$.

The kernel (\ref{k3a}) is invariant relative to rotations of the vectors $\vec{k},\vec{k’}$ around $\vec{n}$. Therefore, choosing for convenience
the $z$-axes along the vector $\vec{n}$ and introducing the  standard spherical angles $\theta,\phi$, we can represent the solutions of Eq. (\ref{k3a}) as
\begin{equation}\label{psi_m}
\psi_{m}(\vec{k},\vec{n})=\psi_{m}(k,\theta)\exp(im\phi).
\end{equation}
Then, after separating the dependence on $\phi$,  the equation (\ref{k3a}) is reduced to the following equation for the function $\psi_{m}(k,\theta)$:
\begin{widetext}
\begin{equation}\label{eqlm}
\left(4(\vec{k}\,^2 + m^2)-M_m^2\right)\psi_{m}(k,\theta)=-\frac{m^2}{2\pi^3} \int  V_m(k,\theta,k’,\theta’,M_m^2) \psi_{m}(k’,\theta’)   
\frac{{k’}^2dk'}{\varepsilon_{k’}}\sin\theta’d\theta’.
 \end{equation}
 \end{widetext}

The kernel $V(\vec{k},\vec{k’},\vec{n};M^2)$ in Eq. (\ref{eqwf3a}) depends on the scalar products
\begin{eqnarray*}
&&\vec{k}\cd\vec{k’}=kk’(\cos\theta\cos\theta’+\sin\theta\sin\theta’ \cos\varphi),
\\
&&\vec{n}\cd\vec{k}=k\cos\theta,\quad \vec{n}\cd\vec{k’}=k’\cos\theta’,
\end{eqnarray*}
that is, $V=V(k,\theta,k’,\theta’,\varphi,M^2)$.
The kernel in Eq. (\ref{eqlm}) is defined as
\begin{equation}\label{Vm}
\! V_m(k,\theta,k’,\theta’,M^2)=\int_0^{2\pi}V(k,\theta,k’,\theta’,\varphi,M^2)\cos(m\varphi) d\varphi.
\end{equation}
Since the kernel $V_m$ depends on the projection $m$, the bound state mass $M_m$ in (\ref{eqlm}) also depends on $m$.
This split of masses $M_m$ vs. $m$ is a manifestation of violation of the rotational invariance. If $\vec{n}$-dependence of the kernel disappears, these energies become degenerate.
The magnitude of this dependence is illustrated numerically in the Tables \ref{tab1} and \ref{tab2} for the mass of the exchanged particle $\mu=0.07$ and $\mu=0.15$ respectively. $B_{\text{NR}}$ and $B_{\text{BS}}$  are the eigenvalues found from the Schr\"odinger and BS equations correspondingly.
\begin{table}[h]
\begin{center}
\caption{Binding energies $B_a$ corresponding to Eq. (\ref{eqwf3a}) with the kernel (\ref{k3a}) for the eigenvalues $a=0,1$ of the operator (\ref{A2}) Case: $L=1$, $m=1$, $\mu=0.07$.}\label{tab1}
\bigskip
\begin{tabular}{l l l l l}\hline
$\alpha$      &  $B_{\text{a=0}}$  &   $B_{\text{a=1}}$ & $B_{\text{BS}}$  & $B_{\text{NR}}$  \\ \hline
$1.28$              & 0.0126    &  0.0086   &  0.01     & 0.034   \cr
  \hline
  $3.59$              & 0.128     &  0.083   &  0.1    &  0.58   \cr
  \hline
\end{tabular}
\end{center}
\end{table}
\noindent
\begin{table}[h]
\begin{center}
\caption{Binding energies $B_a$ corresponding to Eq. (\ref{eqwf3a}) with the kernel (\ref{k3a}) for the eigenvalues $a=0,1$ of the operator (\ref{A2}). Case: $L=1$, $m=1$, $\mu=0.15$.} \label{tab2}
\bigskip
\begin{tabular}{l l l l l}\hline
$\alpha$      &  $B_{\text{a=0}}$  &   $B_{\text{a=1}}$ & $B_{\text{BS}}$  & $B_{\text{NR}}$  \\ \hline
$2.30$              & 0.0183    &  0.0078   &  0.01     & 0.079   \cr
  \hline
  $4.76$              & 0.152     &  0.082   &  0.1    &  0.80   \cr
  \hline
\end{tabular}
\end{center}
\end{table}

The solutions of Eq. (\ref{eqwf3a}) with the kernel (\ref{k3a})  (re-written in slightly different but equivalent form, which will be used below) were found and studied in detail in Ref. \cite{M-J}. 

If we don’t  fix the vector $\vec{n}$ along $z$, but consider it on the equal ground  with other vectors, then the  kernel  $V(\vec{k}\,',\vec{k},\vec{n},M^2)$ is invariant under rotations of the coordinate system.\footnote{Initially, each particle is associated with a four-momentum. Momentum $\vec{k}$ is the spatial part
of the four-vector $k=L^{-1}({\cal P})k_1$, ${\cal P}=k_1+k_2$ and  $L^{-1}({\cal P})$ is the Lorentz transformation to the reference frame where $\vec{\cal P}=0$. The vector $\vec{n}$ is constructed similarly from $\omega$ (and normalized to 1). Under the Lorentz transformations and rotations of all the four-vectors  in the r.h.-side of the formula $k=L^{-1}({\cal P})k_1$, the vector $\vec{k}$ is rotated only (and similarly for $\vec{k’},\vec{n}$, see \cite{cdkm}, Sec. 3.1.2). Therefore, staying within the relativistic framework, we discuss the rotations only of these vectors.} 
Corresponding operator of the infinitesimal rotations is
\begin{equation}\label{Ln}
\vec{L}=\vec{L}_{\vec{k}}+\vec{L}_{\vec{n}}=-i[\vec{k}\times \partial /\partial \vec{k}]-i[\vec{n}\times \partial /\partial \vec{n}],
\end{equation}
the interaction kernel commutes with it. Therefore, solutions of the equation (\ref{eqwf3a}) are also the eigenstates of the operators $\vec{L}^2,L_z$, where 
$\vec{L}$ is defined in (\ref{Ln}).

Note that there exists another operator which commutes not only with the Hamiltonian determining the equation (\ref{eqwf3a}) but also with $\vec{L}$, that is, with $\vec{L}^2,L_z$, namely
\begin{equation}\label{A2}
A^2=(\vec{n}\cd\vec{L})^2.
\end{equation}
The binding energies now depend on $l$, associated with the operator (\ref{Ln}). They don’t depend on its projection $m$ on the $z$-axes but depend on $a$ - the eigenvalues of the operator $A$. That is, when the equation $\omega\cd x=0$ is not associated with any particular coordinate axes, the energies depend on projection of the angular momentum of the direction determining orientation of the LF plane which can be taken arbitrary and can be varied.

As we will discuss below, the dependence on $a$ is a consequence of truncation resulting in the equation (\ref{eqwf3a}) for a separate Fock component - the exact system of equations has solutions with the $a$-independent energies. 
Denote the set of eigenvalues of the operators $\vec{L}^2,L_z$ and $A^2$ as $l(l+1),M,a^2$ and corresponding eigenfunctions as 
$\chi^{(a)}_{lm}(\vec{k},\vec{n})$.
Then the solutions of Eq. (\ref{eqwf3a}) are labeled by  $l,m,a$ and therefore have the form
\begin{equation}\label{Psia}
\Psi^{(a)}_{lm}(\vec{k},\vec{n})=\psi^{(a)}_l(k,\vec{n}\cd\vec{k})\chi^{(a)}_{lm}(\vec{k},\vec{n}),
\end{equation}
where $\chi^{(a)}_{lm}(\vec{k},\vec{n})$ are the common eigenfunctions  of the operators $\vec{L}^2,L_z,A^2$.  
Eq. (\ref{Psia}) is a generalization of the formula (\ref{psi0}) for the case of $\vec{n}$-dependent but scalar kernel.
In the next section we will develop a method of finding the eigenfunctions  $\chi^{(a)}_{lm}(\vec{k},\vec{n})$ and find them explicitly for $l=1,2$.

The states $\Psi^{(a)}_{lm}(\vec{k},\vec{n})$  with $a=0,1,2,\ldots l$ are excessive relative to the solutions with the $\vec{n}$-independent kernel.
For given $l$, which of them should we put in correspondence to the physical system?  The solution was proposed in Refs.\cite{NP2000,NP2001,NP2002} where it was checked that the true binding energies are described fairly well.  In the present work we will also check the wave functions comparing them with the “true” ones.
To find the “true” solutions, we will solve the BS equation and project its solution on the light-front. We will see that the receipt given in \cite{NP2000,NP2001,NP2002} works well for the wave functions too. This breakthrough was achieved now due to progress in finding solutions of the BS equation for the nonzero angular momentum.
\bigskip

\section{Common eigenfunctions of the operators $\vec{L}^2$, $L_z$ and $(\vec{n}\cd\vec{L})^2$}\label{eigenf}

As we established, the solutions of the equation (\ref{eqwf3a}) have the form (\ref{Psia}). They contain the common eigenfunctions 
$\chi^{(a)}_{lm}(\vec{k},\vec{n})$ of the set of operators $\vec{L}^2,L_z,A^2$. In this section we describe the general procedure of finding these functions and 
apply it to the state with $l=1$. 

The operator $\vec{L}$, Eq. (\ref{Ln}), is the sum of two terms $\vec{L}_{\vec{k}}=-i[\vec{k}\times \partial /\partial \vec{k}]$
and $\vec{L}_{\vec{n}}=-i[\vec{n}\times \partial /\partial \vec{n}]$. Therefore, the set of eigenfunctions of the operator $\vec{L}$ is constructed by the well known formulas for addition of the angular momenta, namely:
\begin{equation}\label{chiLm}
\chi_{lm}^{l_1l_2}(\hat{\vec{k}},\vec{n})=\sum_{m_1m_2}C_{l_1m_1l_2m_2}^{lm}Y_{l_1m_1}(\hat{\vec{k}})Y_{l_2m_2}(\vec{n}),
\end{equation}
where $C_{l_1m_1l_2m_2}^{lm}$ are the Clebsch-Gordan coefficients, $\hat{\vec{k}}=\vec{k}/k$.
In (\ref{chiLm}) we take $l_1+l_2=l$ since the functions with $l_1+l_2>l$ are expressed via the functions (\ref{chiLm}).
$l_1$ can run through  the values $l_1=0,1,2,\ldots, l$ (and similarly for $l_2=l-l_1$).
The number of elements in the set (\ref{chiLm}) is the same as the number of the eigenfunctions with $a=0,1,2,\ldots,l$.
We also should find the eigenfunctions of the operator (\ref{A2}):
\begin{equation}\label{A2eig}
A^2\chi^{(a)}(\vec{k},\vec{n})=a^2 \chi^{(a)}(\vec{k},\vec{n}), \quad a^2=0,1,2^2, \ldots,l^2.
\end{equation}
Since the functions $\chi^{(a)}(\vec{k},\vec{n})$ should be also the eigenfunctions of the operators $\vec{L}^2,L_z$ (the latter ones are provided by Eq. (\ref{chiLm})),  $\chi^{(a)}(\vec{k},\vec{n})$ should be superpositions of the functions (\ref{chiLm}):
 \begin{eqnarray}\label{chiLma}
\chi^{(a)}_{lm}(\vec{k},\vec{n})&=&\sum_{l_1+l_2=l}h^{(a)}_{l_1l_2}
\sum_{m_1+m_2=m}C_{l_1m_1l_2m_2}^{lm}
\nonumber\\
&\times&Y_{l_1m_1}(\hat{\vec{k}})Y_{l_2m_2}(\vec{n}),
 \end{eqnarray}
 with the coefficients $h^{(a)}_{l_1l_2}$ found from the equation (\ref{A2eig}). 
 Note that $(\vec{n}\cd\vec{L})f(\vec{n}\cd \vec{k})=0$, where $f(\vec{n}\cd \vec{k})$ is arbitrary function
since $(\vec{n}\cd \vec{L})$ is the operator of the infinitesimal rotations around the vector $\vec{n}$ which don't change the angle between $\vec{k}$ and $\vec{n}$ and, hence, don't change the scalar product  $(\vec{n}\cd \vec{k})$. Therefore the coefficients $h^{(a)}_{l_1l_2}$ can depend on  
$\vec{n}\cd \vec{k}$.

Below we will denote the correlated pair $l_1,l_2$ as $j$: 
\begin{widetext}
$$
j=[1,2,\ldots,l+1]=[(l_1=0,l_2=l),(l_1=1,l_2=l-1), \ldots,(l_1=l,l_2=0)]
$$
\end{widetext} 
and $m+1$ (for $m\ge 0$) as $i$: $i=m+1=1,2,\ldots,l+1$. Then Eq. (\ref{chiLma}) is rewritten as:
$$
\chi_i=\sum_{j=1}^{l+1}\mathcal{A}_{ij}h_j,
$$
where $\mathcal{A}_{ij}$ is determined by (\ref{chiLma}).
 Substituting in Eq. (\ref{A2eig}) the function $\chi^{(a)}_{lm}(\vec{k},\vec{n})$, Eq. (\ref{chiLma}), we represent the system of equations in the form:
 \begin{equation}\label{eqL3a}
\sum_{j=1}^{l+1}\mathcal{B}_{ij}h_j=a^2\sum_{j=1}^{l+1}\mathcal{A}_{ij}h_j,
\end{equation}
where $\mathcal{B}_{ij},\;\mathcal{A}_{ij}$ are the $(l+1)\times(l+1)$ square matrices:
\begin{eqnarray*}
\mathcal{B}_{ij}&=&\sum_{m_1+m_2=m} C_{l_1m_1l_2m_2}^{lm}
\Bigl\{(\vec{n}\cd\vec{L})^2Y_{l_1m_1}(\hat{\vec{k}})\Bigr\}Y_{l_2m_2}(\vec{n}),
\\
\mathcal{A}_{ij}&=&\sum_{m_1+m_2=m} C_{l_1m_1l_2m_2}^{lm} Y_{l_1m_1}(\hat{\vec{k}})Y_{l_2m_2}(\vec{n}).
\end{eqnarray*}
In this way, we obtain the eigenstate equation:
\begin{equation}\label{eqL3b}
Hh=\lambda h,
\end{equation}
where $H =\mathcal{A}^{-1}\mathcal{B}$ and  $\lambda= a^2$.
 
Due to the parity conservation the values $\pm m$ provide the same equations, therefore it is enough to take the values $m\ge 0$. We also used that the operator $(\vec{n}\cd\vec{L})^2$ applied to the product 
$Y_{l_1m_1}(\hat{\vec{k}})Y_{l_2m_2}(\vec{n})$ acts on $Y_{l_1m_1}(\hat{\vec{k}})$ only.
In the spherical coordinates the components of  $\hat{\vec{k}},\,\vec{n}$ read: 
\begin{eqnarray*}
&&\hat{k}_x=\sin\theta_k\cos\phi_k,\;\hat{k}_y=\sin\theta_k\sin\phi_k,\;\hat{k}_z=\cos\theta_k,
\\
&&n_x=\sin\theta_n \cos\phi_n,\;n_y=\sin\theta_n \sin\phi_n,\;n_z=\cos\theta_n.
\end{eqnarray*}
The scalar product $\vec{n}\cd\hat{\vec{k}}$ obtains the form:
\begin{equation}\label{nk}
\vec{n}\cd\hat{\vec{k}}=\cos\theta_k\cos\theta_n+\cos(\phi_k-\phi_n)\sin\theta_k\sin\theta_n.
\end{equation}

We introduce also the components $L_{\pm}=L_x\pm iL_y$, $n_{\pm}=n_x\pm in_y=\sin\theta_n \exp(\pm i\phi_n)$. In the spherical coordinates the operator
$$
\vec{L}_{\vec{k}}=-i[\vec{k}\times\partial/ \partial\vec{k}]
$$
obtains the form
$$
L_{\pm}=\exp(\pm i \phi_k)\left(\pm\frac{\partial}{\partial \theta_k}+i\cot\theta_k \frac{\partial}{\partial \phi_k}\right),\quad L_z=-i\frac{\partial}{\partial \phi_k}.
$$
Then the scalar product $(\vec{n}\cd\vec{L})$ reads:
\begin{eqnarray}\label{nL}
(\vec{n}\cd\vec{L})&=&i\sin(\phi_k-\phi_n)\sin\theta_n\frac{\partial}{\partial \theta_k}
\\
&-&i\Bigl[\cos\theta_n-\cos(\phi_k-\phi_n)\cot\theta_k\sin\theta_n\Bigr]
\frac{\partial}{\partial \phi_k}.
\nonumber
\end{eqnarray}

Below, for $l=1$ and $l=2$, we obtain the explicit form of  the eigenvalue equations (\ref{eqL3b}). Solving it,  we find the eigenfunctions (\ref{chiLma}).
\bigskip

\subsection{The $l=1$ state}\label{L1}
Let us consider the case $l=1$. Equation (\ref{chiLma}) for $m=0,1$ obtains the form:
\begin{eqnarray}\label{AL1}
&& \chi^{(a)}_{10}(\vec{k},\vec{n})=h_{10}\sqrt{3}\cos\theta_k+ h_{01}\sqrt{3}\cos\theta_n
\\
&&  \chi^{(a)}_{11}(\vec{k},\vec{n})=-h_{10}\sqrt{\frac{3}{2}}e^{i\phi_k}\sin\theta_k 
-h_{01}\sqrt{\frac{3}{2}}e^{i\phi_n}\sin\theta_n
\nonumber
\end{eqnarray}
In this way, we find the matrix $\mathcal{A}$:
\begin{equation}\label{A1tab}
\mathcal{A}=
\left(
\begin{array}{cc}
\sqrt{3}\cos\theta_k & \sqrt{3}\cos\theta_n
\\
-\sqrt{\frac{3}{2}}e^{i\phi_k}\sin\theta_k & -\sqrt{\frac{3}{2}}e^{i\phi_n}\sin\theta_n
\end{array}
\right)
\end{equation}
Applying to Eq. (\ref{AL1}) twice the operator $(\vec{n}\cd\vec{L})$, defined in (\ref{nL}), we get:
\begin{eqnarray}\label{BL1}
(\vec{n}\cd\vec{L})^2\chi^{(a)}_{10}(\vec{k},\vec{n})&=&h_{10}\sqrt{3}\sin\theta_n[\cos\theta_k\sin\theta_n
\nonumber\\
&-&\cos(\phi_k-\phi_n)\cos\theta_n\sin\theta_k]
\nonumber\\
(\vec{n}\cd\vec{L})^2\chi^{(a)}_{11}(\vec{k},\vec{n})
&=&-h_{10}\sqrt{\frac{3}{32}}e^{-i\phi_k}\{\sin\theta_k[e^{2i\phi_k}(3+\cos 2\theta_n)
\nonumber\\
&-&2e^{2i\phi_n}\sin^2\theta_n]
\nonumber\\
&-&2e^{i(\phi_k+\phi_n)}\cos\theta_k\sin 2\theta_n\}
\end{eqnarray}
In this way, we find the elements of the matrix $\mathcal{B}$:
\begin{eqnarray*}
\mathcal{B}_{11}&=&\sqrt{3}\sin\theta_n[\cos\theta_k\sin\theta_n-\cos(\phi_k-\phi_n)\cos\theta_n\sin\theta_k]
\\
\mathcal{B}_{12}&=&0
\\
\mathcal{B}_{21}&=&-\frac{1}{4}\sqrt{\frac{3}{2}}e^{-i\phi_k}\{\sin\theta_k[e^{2i\phi_k}(3+\cos(2\theta_n))
\\
&-&
2e^{2i\phi_n}\sin^2\theta_n]
-2e^{i(\phi_k+\phi_n)}\cos\theta_k\sin(2\theta_n)\}
\\
\mathcal{B}_{22}&=&0
\end{eqnarray*}

Matrix $H=\mathcal{A}^{-1}\mathcal{B}$:
\begin{equation}\label{H1tab}
H=
\left(
\begin{array}{cc}
1&0
\\
-\vec{n}\cd\hat{\vec{k}}  &0
\end{array}
\right)
\end{equation}
Eigensystem:
$$
h_0=
\left(
\begin{array}{c}
0
\\
1
\end{array}
\right),\;
h_1=
\left(
\begin{array}{c}
-1 
\\
\vec{n}\cd\hat{\vec{k}}
\end{array}
\right).
$$
Then, by Eq. (\ref{chiLma})
we find the eigenfunctions $\chi_{1m}^{(a)}(\vec{k},\vec{n})$:
\begin{eqnarray}\label{chiL1}
 \chi_{1m}^{(0)}(\vec{k},\vec{n})&=&(\vec{n}\cd\hat{\vec{k}})Y_{1m}(\vec{n}),
 \nonumber\\
 \chi_{1m}^{(1)}(\vec{k},\vec{n})&=&Y_{1m}(\hat{\vec{k}})- (\vec{n}\cd\hat{\vec{k}})Y_{1m}(\vec{n}),
\end{eqnarray}
 $\vec{n}\cd\hat{\vec{k}}$ is defined in (\ref{nk}).
They are orthogonal.  Namely (without integration):
\begin{equation}\label{norm3a}
         \sum_m\chi^{(0)*}_{1m}(\hat{\vec{k}},\vec{n})\chi^{(1)}_{1m}(\hat{\vec{k’}},\vec{n})=0.
\end{equation}
Note also:
 \begin{eqnarray}\label{norm3b}
           && \sum_m\chi^{(0)*}_{1m}(\hat{\vec{k}},\vec{n})\chi^{(0)}_{1m}(\hat{\vec{k’}},\vec{n})=\frac{3}{4\pi}zz’,
               \nonumber\\   
           && \sum_m\chi^{(1)*}_{1m}(\hat{\vec{k}},\vec{n})\chi^{(1)}_{1m}(\hat{\vec{k’}},\vec{n})=
        \frac{3}{4\pi}[\hat{\vec{k}}\cd\hat{\vec{k’}}-zz’],
\end{eqnarray}
where $z=\vec{n}\cd\hat{\vec{k}}$ and $z'=\vec{n}\cd\hat{\vec{k'}}$.
Except for normalization,  these eigenfunctions coincide with ones found in   Refs. \cite{NP2000,NP2001}. 

\subsection{The $l=2$ state}\label{L2}

Let us consider the case $l=2$:
\begin{eqnarray}\label{chiL2m}
\chi^{(a)}_{2m}(\vec{k},\vec{n})&=&h_{20}Y_{2m}(\hat{\vec{k}})Y_{00}
\nonumber\\
&+&h_{11}\sum_{m_1+m_2=m}C_{1m_11m_2}^{2m}Y_{1m_1}(\hat{\vec{k}})Y_{1m_2}(\vec{n})
\nonumber\\
&+&h_{02}Y_{00}Y_{2m}(\vec{n}).
\end{eqnarray}
Similarly to Eq. (\ref{AL1}), for $m=0,1,2$, we take from (\ref{chiL2m}) the coefficients at $h_{20},h_{11}, h_{02}$ and fill the  $3\times 3$ matrix $\mathcal{A}$.

Similarly to Eqs. (\ref{BL1}), we calculate $(\vec{n}\cd \vec{L})^2\chi^{(a)}_{2m}(\vec{k},\vec{n})$ and fill the  $3\times 3$ matrix $\mathcal{B}$. Then we find the $3\times 3$ matrix $H=\mathcal{A}^{-1}\mathcal{B}$. It reads:
\begin{equation}\label{H2tab}
H=\mathcal{A}^{-1}\mathcal{B}=
\left(
\begin{array}{ccc}
0 & -\sqrt{\frac{6}{5}}(\vec{n}\cd\hat{\vec{k}}) & 2
\\
0&1& -\sqrt{30}(\vec{n}\cd\hat{\vec{k}})
\\
0 & 0 & 4
\end{array}
\right)
\end{equation}
\bigskip

Corresponding eigensystem, solution of the equation $Hh_a=a^2h_a$:
\begin{eqnarray*}
h_0&=&
\left(\begin{array}{c}
1
\\
0
\\
0
\end{array}
\right),\;
h_1=
\left(\begin{array}{c}
-\sqrt{\frac{6}{5}}(\vec{n}\cd\hat{\vec{k}})
\\
1
\\
0
\end{array}
\right),
\\
h_2&=&
\left(\begin{array}{c}
\frac{1}{2}(1+(\vec{n}\cd\hat{\vec{k}})^2)
\\
-\sqrt{\frac{10}{3}}(\vec{n}\cd\hat{\vec{k}})
\\
1
\end{array}
\right)
\end{eqnarray*}
\normalsize

Multiplying them at the right by the raw 
\begin{eqnarray*}
Y&=&
\Bigl(Y_{00}Y_{2m}(\vec{n}),\;
\sum
C_{1m_11m_2}^{2m}Y_{1m_1}(\hat{\vec{k}})Y_{1m_2}(\vec{n}),
\\
&&Y_{2m}(\hat{\vec{k}})Y_{00}\Bigr)
\end{eqnarray*}
\normalsize
we get the wave functions:
\footnotesize
\begin{eqnarray}\label{chiL2}
\chi^{(0)}_{2m}(\vec{k},\vec{n})&=&Y_{00}Y_{2m}(\vec{n}),
\nonumber\\
\chi^{(1)}_{2m}(\vec{k},\vec{n})&=&\sum_{m_1+m_2=m}C_{1m_11m_2}^{2m}Y_{1m_1}(\hat{\vec{k}})Y_{1m_2}(\vec{n})
\nonumber\\
&-&\sqrt{\frac{6}{5}}(\vec{n}\cd\hat{\vec{k}})Y_{00}Y_{2m}(\vec{n}),
\nonumber\\
\chi^{(2)}_{2m}(\vec{k},\vec{n})&=&Y_{2m}(\hat{\vec{k}})Y_{00}
\nonumber\\
&-&\sqrt{\frac{10}{3}}(\vec{n}\cd\hat{\vec{k}})\sum_{m_1+m_2=m}C_{1m_11m_2}^{2m}Y_{1m_1}(\hat{\vec{k}})Y_{1m_2}(\vec{n})
\nonumber\\
&+&\frac{1}{2}[1+(\vec{n}\cd\hat{\vec{k}})^2]Y_{2m}(\vec{n})Y_{00}.
\end{eqnarray}
\normalsize
 On can check by the direct substitution that the eigenfunctions (\ref{chiL2}) indeed satisfy the equation (\ref{A2eig}) with $l=2$.
 
 These functions are orthogonal:
 \begin{widetext}
 \begin{equation}\label{norm4a}
 \sum_m\chi^{(0)*}_{2m}(\vec{k},\vec{n})\chi^{(1)}_{2m}(\vec{k’},\vec{n})=\sum_m\chi^{(0)*}_{2m}(\vec{k},\vec{n})\chi^{(2)}_{2m}(\vec{k’},\vec{n})
=\sum_m\chi^{(1)*}_{2m}(\vec{k},\vec{n})\chi^{(2)}_{2m}(\vec{k’},\vec{n})=0.
\end{equation}
 Note that
\begin{eqnarray}\label{norm4b}
&&\sum_m\chi^{(0)*}_{2m}(\vec{k},\vec{n})\chi^{(0)}_{2m}(\vec{k’},\vec{n})=\frac{5}{(4\pi)^2} ,\quad 
\sum_m\chi^{(1)*}_{2m}(\vec{k},\vec{n})\chi^{(1)}_{2m}(\vec{k’},\vec{n})
=\frac{9}{2(4\pi)^2}[\hat{\vec{k}}\cd\hat{\vec{k’}}-zz’],
\nonumber\\
&& \sum_m\chi^{(2)*}_{2m}(\vec{k},\vec{n})\chi^{(2)}_{2m}(\vec{k’},\vec{n})=\frac{15}{4(4\pi)^2}[z^2+{z’}^2+
z^2{z’}^2+2(\hat{\vec{k}}\cd\hat{\vec{k’}})^2-
4zz’(\hat{\vec{k}}\cd\hat{\vec{k’}})-1],
\end{eqnarray}
\end{widetext}
$z,z’$ are defined below Eq. (\ref{norm3b}).

 \section{Bound state solutions}\label{bound1}
 \subsection{$l=1$}\label{bsL1}
 
 For the eigenvalues $l=1,m=0,\pm 1$ and $a=0,1$ of the operators $\vec{L}^2,L_z$ and $A^2=(\vec{n}\cd\vec{L})^2$, we are looking for solution of Eq. (\ref{eqwf3a}) in the form 
 \begin{eqnarray}\label{bspsi1}
 \psi^{(0)}_{1m}(\vec{k},\vec{n})&=&c_0 g_1^{(0)}(k,\theta)\chi_{1m}^{(0)}(\vec{k},\vec{n}),
\nonumber\\
\psi^{(1)}_{1m}(\vec{k},\vec{n})&=&c_1 g_1^{(1)}(k,\theta)\chi_{1m}^{(1)}(\vec{k},\vec{n}),
\end{eqnarray}
with $\chi_{1m}^{(0,1)}(\vec{k},\vec{n})$ defined in (\ref{chiL1}) and $\cos \theta=\vec{n}\cd\hat{\vec{k}}=z$.

We substitute the wave functions  (\ref{bspsi1}) into Eq. (\ref{eqwf3a}), multiply it in turn by $\chi^{(0)*}_{1m}(\hat{\vec{k}},\vec{n})$, 
 $\chi^{(1)*}_{1m}(\hat{\vec{k}},\vec{n})$, take sum over $m=0,\pm 1$ and use the formulas (\ref{norm3a},\ref{norm3b}).
Then from Eq. (\ref{eqwf3a}) we obtain equations for $g^{(0,1)}_1$
\begin{widetext}
\begin{equation}\label{eq10}
\left(4(\vec{k}\,^2 + m^2)-M_0^2\right)[ g^{(0)}_1(k,\theta)\cos\theta] =
-\frac{m^2}{2\pi^3} \int  V_0(k,\theta,k’,\theta’,M_0^2) [g^{(0)}_1(k’,\theta’)\cos\theta’]\sin\theta’d\theta’
 \frac{{k’}^2 dk'}{\varepsilon_{k’}},
\end{equation}
\begin{equation}\label{eq11}
\left(4(\vec{k}\,^2 + m^2)-M_1^2\right)[ g^{(1)}_1(k,\theta)\sin\theta] =
-\frac{m^2}{2\pi^3} \int  V_1(k,\theta,k’,\theta’,M_1^2)[ g^{(1)}_1(k’,\theta’)\sin\theta’]\sin\theta’d\theta’
 \frac{{k’}^2 dk'}{\varepsilon_{k’}}.
\end{equation}
\end{widetext}
Denoting 
\begin{eqnarray*}
\psi_1^{(0)}(k,\theta)&=&g^{(0)}_1(k,\theta)\cos\theta,
\\
\psi_1^{(1)}(k,\theta)&=&g^{(1)}_1(k,\theta)\sin\theta,
\end{eqnarray*}
we obtain equation for $\psi_1^{(a)}(k,\theta)$
\begin{widetext}
\begin{equation}\label{eqL1a}
\left(4(\vec{k}\,^2 + m^2)-M_a^2\right)\psi^{(a)}_1(k,\theta)=-\frac{m^2}{2\pi^3} \int  V_a(k,\theta,k’,\theta’,M_a^2) \psi^{(a)}_1(k’,\theta’)   
\frac{{k’}^2dk'}{\varepsilon_{k’}}\sin\theta’d\theta’.
 \end{equation}
 \end{widetext}
The kernel $V(\vec{k},\vec{k’},\vec{n};M^2)$ in Eq. (\ref{eqwf3a}) depends on the scalar products
\begin{eqnarray*}
&&\vec{k}\cd\vec{k’}=kk’(\cos\theta\cos\theta’+\sin\theta\sin\theta’ \cos\varphi),
\\
&&\vec{n}\cd\vec{k}=k\cos\theta,\quad \vec{n}\cd\vec{k’}=k’\cos\theta’,
\end{eqnarray*}
that is, $V=V(k,\theta,k’,\theta’,\varphi,M^2)$.
The kernels in Eqs. (\ref{eq10}), (\ref{eq11}) and (\ref{eqL1a}) are defined as
\begin{equation}\label{Va}
V_a(k,\theta,k’,\theta’,M^2)=\int_0^{2\pi}V(k,\theta,k’,\theta’,\varphi,M^2)\cos(a\varphi) d\varphi
\end{equation}
with $a=0,1$. If the kernel $V=V(k,\theta,k’,\theta’,\varphi,M^2)$ depends on $\vec{n}$, the equations (\ref{eq10}) and (\ref{eq11}) determine
different energies. This is a manifestation of violation of the rotational invariance. If its $\vec{n}$-dependence disappears, these energies become equal to each other. The latter follows from the fact that both equations 
(\ref{eq10}) and (\ref{eq11})  are consequences of Eq. (\ref{eqwf3a}). For the $\vec{n}$-independent kernel the energies determined by Eq.  (\ref{eqwf3a}) 
don’t depend on projection of the angular momentum on any direction.  Therefore, split of energies found from Eqs.  (\ref{eq10}), (\ref{eq11}), like the 
$\vec{n}$-dependence of the kernel, is a relativistic effect and, hence, it is small in systems with small binding energies and momenta. As it was shown in Sec. \ref{equation}, in this limit the kernel (\ref{k3a}) turns into the Yukawa one, Eq. (\ref{k5}).

Note that after identifying $a$ with $m$, the equation (\ref{eqL1a}) coincides with (\ref{eqlm}). However, because of violation of the rotational symmetry, 
the solutions in Eq. (\ref{eqlm}) are not associated with any angular momentum. The solutions of Eq. (\ref{eqL1a}) are associated with the angular momentum
(in particular, $l=1$ in Eq. (\ref{eqL1a})), defined by means of the operator (\ref{Ln}). This property remains to be valid for higher angular momenta.

 \subsection{$l=2$}\label{bsL2}
 
Similarly to Eq. (\ref{bspsi1}),  we are again looking for solution in the form:
\begin{equation}\label{psi2a}
\psi_{2m}^{(a)}(\vec{k},\vec{n})=g_2^{(a)}(k,z)C^{(a)}(z)\chi_{2m}^{(a)}(\vec{k},\vec{n}),
\end{equation}
where the functions  $\chi_{2m}^{(a)}(\vec{k},\vec{n}),\; a=0,1,2,$ are defined in (\ref{chiL2}). For the reasons which will be explained below (in Sec. \ref{psi_2}), 
we introduced in (\ref{psi2a}) the coefficients  
 \begin{equation}\label{C012}
C^{(0)}=\frac{1}{2}(-1+3z^2),\; C^{(1)}=\sqrt{\frac{10}{3}}z,\; C^{(2)}=1.
\end{equation} 
Since these coefficients are the scalars, the operator $\vec{L}$ acting on them gives zero (and, hence, all three operators $\vec{L}^2,L_z, A^2$ also give zero). Therefore the products $C_a(z)\chi_{2m}^{(a)}(\vec{k},\vec{n})$ are the eigenfunctions of these operators, like 
$\chi_{2m}^{(a)}(\vec{k},\vec{n})$.

We substitute the wave function  (\ref{psi2a}) into Eq. (\ref{eqwf3a}), 
multiply it in turn by $\chi^{(0,1,2)*}_{2m}(\vec{k},\vec{n})$, 
and take sum over $m=0,\pm 1,\pm 2$. We use the properties (\ref{norm4a}), (\ref{norm4b}). Due to orthogonality,  Eq.  (\ref{norm4a}), we obtain the uncoupled equations for the scalar functions $g_2^{(a)}$:
\begin{widetext}
\begin{eqnarray}\label{eqL20b}
\left(4(\vec{k}\,^2 + m^2)-M_0^2\right)(-1+3z^2)g^{(0)}_2(k,z)&=&-\frac{m^2}{2\pi^3} \int  V(\vec{k},\vec{k}\,’,\vec{n},M_0^2)(-1+3{z’}^2) g^{(0)}_2(k’,z’)  \frac{d^3k'}{\varepsilon_{k'}},
 \\
\label{eqL21b}
\left(4(\vec{k}\,^2 + m^2)-M_1^2\right)(1-z^2)zg^{(1)}_2(k,z) &=&
-\frac{m^2}{2\pi^3} \int  V(\vec{k},\vec{k}\,’,\vec{n},M_1^2)(\hat{\vec{k}}\cd\hat{\vec{k’}}-zz’)z’g^{(1)}_2(k’,z’) 
 \frac{d^3k'}{\varepsilon_{k'}},
 \nonumber\\
 &&
 \\
\left(4(\vec{k}\,^2 + m^2)-M_2^2\right)(1-z^2)^2g^{(2)}_2(k,z)&=&
-\frac{m^2}{2\pi^3} \int  V(\vec{k},\vec{k}\,’,\vec{n},M_2^2)
\nonumber\\
&\times& 
[z^2+{z’}^2+z^2{z’}^2+2(\hat{\vec{k}}\cd\hat{\vec{k’}})^2-4zz’(\hat{\vec{k}}\cd\hat{\vec{k’}})-1]g^{(2)}_2(k’,z’) 
 \frac{d^3k'}{\varepsilon_{k'}}.
 \label{eqL22b}
\end{eqnarray}
We remind that $z=\vec{n}\cd\vec{k}/k$, $z’=\vec{n}\cd\vec{k’}/k’$. 
These equations determine different eigenvalues $M_a$.
In terms of the angular variables:
\begin{eqnarray}
\left(4(\vec{k}\,^2 + m^2)-M_0^2\right)(-1+3\cos^2\theta)g^{(0)}_2(k,\cos\theta)&=&-\frac{m^2}{2\pi^3} \int  V_0(k,\theta,k’,\theta’,M_0^2)(-1+3\cos^2\theta’) 
g^{(0)}_2(k’,\cos\theta’) \sin\theta’d\theta’ \frac{{k’}^2 dk'}{\varepsilon_{k’}},
\nonumber\\
&&\label{eqL20c}
\\
\left(4(\vec{k}\,^2 + m^2)-M_1^2\right)\sin\theta\cos\theta\,g^{(1)}_2(k,\cos\theta) &=&
-\frac{m^2}{2\pi^3} \int  V_1(k,\theta,k’,\theta’,M_1^2)\sin\theta’\cos\theta’\,g^{(1)}_2(k’,\cos\theta’) 
\sin\theta’d\theta’ \frac{{k’}^2 dk'}{\varepsilon_{k’}},
\label{eqL21c}
\\
\left(4(\vec{k}\,^2 + m^2)-M_2^2\right)\sin^2\theta\,g^{(2)}_2(k,\cos\theta) &=&
-\frac{m^2}{2\pi^3} \int  V_2(k,\theta,k’,\theta’,M_2^2)\sin^2\theta’\,g^{(2)}_2(k’,\cos\theta’) 
\sin\theta’d\theta’ \frac{{k’}^2 dk'}{\varepsilon_{k’}},
 \label{eqL22c}
\end{eqnarray}
\end{widetext}
with $V_{a}(k,\theta,k’,\theta’,M_{a}^2)$ (here $a=0,1,2$) defined in (\ref{Va}).

Denoting
\begin{eqnarray}\label{psi012}
\psi^{(0)}_2(k,\theta)&=&(-1+3\cos^2\theta)\,g^{(0)}_2(k,\cos\theta),
\nonumber\\
\psi^{(1)}_2(k,\theta)&=&\sin\theta\cos\theta\, g^{(1)}_2(k,\cos\theta),
\nonumber\\
 \psi^{(2)}_2(k,\theta)&=&\sin^2\theta\, g^{(2)}_2(k,\cos\theta),
\end{eqnarray}
we obtain equation for $ \psi^{(a)}_2(k,\theta)$ the same equation (\ref{eqL1a}) as for $ \psi^{(a)}_1(k,\theta)$, though $a$ can run now until $a=2$. This equation also coincides with Eq. (\ref{eqlm}) after changing notations $a\to m$. 

We emphasize: we could find the spectrum determined by the equation (\ref{eqwf3a}) and the corresponding wave functions $\psi_m$, directly, 
representing the latter ones by Eq. (\ref{psi_m}). Using here more sophisticated method, representing the wave functions via $\chi^{(a)}_{lm}$ by Eq. (\ref{Psia}), we, of course, obtain the same spectrum as in the Eq. (\ref{eqlm}),  however, different classification of the solutions, allowing to distribute the wave functions into the groups with given fixed angular momenta $l$ determined by the operator (\ref{Ln}) and with different $a=0,1,\ldots, l$. As we will see below, physical solutions are superpositions of the states belonging only to a group with fixed $l$. The states from the groups with different angular momenta are not mixed with each other in one superposition.

\section{Finding the physical solution}\label{ph-sol}

We have demonstrated that solutions of the equation (\ref{eqwf3a}) and corresponding eigenvalues are labeled, for given $l,m$, by the addition quantum number $a^2$ -- the eigenvalue of the operator  $A^2=(\vec{n}\cd\vec{L})^2$. This is a consequence of existence of set of the operators $\vec{L}^2$,  $L_z$, $A^2$ which commute with each other.  This is a property of full field-theoretical state vector and, hence, it takes place in any Fock sector. In practice, the corresponding masses $M_a$ are different for different $a$. As mentioned, this split of masses appears because of truncation and depends on it.

It is hardly possible to solve  without truncation
 the exact field-theoretical problem, containing infinite number
of degrees of freedom and
keeping the infinite number of Fock sectors. However, based on the commutation relations,  it is possible to prove the following two important general statements concerning properties of this solution:

 ({\it i}) in the exact solution, all the masses $M_a$ vs. $a$ are the same - they are degenerate;

({\it ii}) the state vector $\Phi$ must have the form of superposition of the degenerate states $\Phi_a$: 
$\Phi=\sum_a c_a\Phi_a$. 

The reason of the property ({\it ii}) is the following:
besides the eigenvalue equation, the state vector satisfies another constraint sometimes called the “angular condition” which so far has not been taken into account. The superposition $\Phi=\sum_a c_a\Phi_a$ satisfies the angular condition.

These statements were proved in Ref. \cite{karmanov82} and reviewed in \cite{cdkm}. For self-sufficiency of presentation, we will briefly repeat their proof here.  

The state vector is defined, in general, on a space-like hypersurface (not necessary plane) in the 4D Minkowski space. 
Under variation of this hypersurface around the point $x$, leaving the hypersurface still space-like one, the variation of the state vector is described by the Tomonaga-Schwinger equation
\cite{Schw}:
\begin{equation}\label{TS}
i\delta\phi/\delta\sigma(x)=H^{int}(x)\phi.
\end{equation}
From this definition, we obtain:
\begin{equation}\label{dphi}
i\delta\phi=H^{int}(x)\phi\delta V(x),
\end{equation}
where $\delta V(x)$ is the volume between the initial surface and the surface obtained from the initial one by the variation $\delta\sigma(x)$ around the point $x$. 

Variations corresponding to translation of the hypersurface as a whole result in the  Schr\"odinger equation in the interaction representation. Its perturbative solution  results in the Feynman graph techniques. 

Making variations corresponding to rotation of the LF plane, that is: $\omega_{\mu}\to \omega’_{\mu}=\omega_{\mu}+\delta\omega_{\mu}$ with 
$\delta\omega_{\mu}=\varepsilon_{\nu\mu}\omega^{\nu}$ ($\epsilon_{\nu\mu}$ is an antisymmetric tensor), we get:
\begin{eqnarray*}                                                                          
\phi_\omega &\rightarrow& \phi_{\omega+\delta\omega} =\phi_\omega                       
+\delta\phi_\omega,
\\
\delta\phi_\omega&=&\frac{1}{2}\epsilon_{\mu\nu}                          
\left( \omega^{\mu}\frac{\partial}{\partial\omega_{\nu}} 
-\omega^{\nu}          
\frac{\partial}{\partial\omega_{\mu}}\right)\phi_\omega\ .                     
\end{eqnarray*}  

The increment of the volume over the point $x$ is:                              
\begin{equation}                                                                               
\delta V =\epsilon_{\mu\nu} x^{\mu} \omega^{\nu} \delta(\omega\cd x) d^4x=
        \frac{1}{2} \epsilon_{\mu\nu} (x^{\mu} \omega^{\nu}- x^{\nu} \omega^{\mu}) \delta(\omega\cd x) d^4x.                                                
\end{equation}     
In this way, we obtain from (\ref{dphi}):                                                                                                    
\begin{equation}\label{kt12}           
L_{\mu\nu}(\omega)\phi_\omega
  =                                       
\hat{J}^{int}_{\mu\nu}            
\ \phi_\omega \ ,                                                               
\end{equation}                                                                  
where:                                                                          
\begin{equation}\label{kt13}                                                    
\hat{L}_{\mu\nu}(\omega) =i\left(\omega_{\mu}                                   
\frac{\partial}{\partial\omega^{\nu}} -\omega_{\nu}                             
\frac{\partial}{\partial\omega^{\mu}}\right)                           
\end{equation}                                                                  
and 
\begin{equation}\label{kt2}                                                     
\hat{J}^{int}_{\mu\nu}=\int H^{int}(x)(x_{\mu}\omega_{\nu} -x_{\nu}\omega_{\mu}) \delta(\omega\cd x)\ d^4x\ .            
\end{equation}
 
 Let us define the operator
 $$
 \Delta \hat{J}_{\mu\nu}   =  \hat{L}_{\mu\nu} -\hat{J}^{int}_{\mu\nu}.
 $$ 
Since the commutators of $\hat{L}_{\mu\nu}$ and $\hat{J}^{int}_{\mu\nu}$ with the four-momentum operator $\hat{P}_{\mu}$, are equal to each other
\cite{cdkm}, the commutator of the operator  $\Delta \hat{J}_{\mu\nu}$ with $\hat{P}_{\mu}$ and, hence, with the mass operator $\hat{P}^2$, is zero. However, 
$[\Delta \hat{J}_{\mu\nu},A]\neq 0$. This means that the state $\Phi’=\Delta \hat{J}_{\mu\nu}\Phi_a$ is not anymore an eigenstate of the operator $A$,
that is, it can be represented in the form $\Phi’=\sum_a c_a\Phi_a$, however, because of $[\Delta \hat{J}_{\mu\nu},P^2]= 0$,
 it belongs to the same mass $M_a$ as $\Phi_a$. This means that, for given $l$, all the states 
$\Phi_a$ correspond to the same mass, that is,  ({\it i}) they are degenerate. ({\it ii}) The state vector $\Phi’$ satisfying the angular condition (\ref{kt12}) has the form of superposition $\Phi’=\sum_a c_a\Phi_a$.  The equation $\Delta \hat{J}_{\mu\nu}\Phi’=0$ determines the coefficients $c_a$ in this superposition.

The superposition takes place in any Fock sector. 
After truncation, keeping, for example, the two-body sector only, we still get solution in the form of superposition of the two-body wave functions with different $a$’s.  However, the states in this superposition, satisfying, e.g., the two-body equation, cease to be degenerate, the solution in the form of superposition 
contains sum of the states with different masses $M_a$. That is, it becomes to be approximate.
This is a natural result of truncation -- result of replacement of the exact Hamiltonian by the approximate (truncated) one.

Though the condition $\Delta \hat{J}_{\mu\nu}\Phi’=0$ requires that the physical solution $\Phi’$ is a superposition of the states 
$\Phi_a$ and, in principle, determines the coefficients in this superposition, for the present, it can be hardly used in practice since the technics allowing to calculate the action of the operator $\hat{J}_{\mu\nu}$, Eq. (\ref{kt2}), on the state vector is not developed. Therefore, instead of Eq. (\ref{kt12}) we will use other condition which seems natural and more applicable. Namely, we suppose (and check below) that the Fock component can depend on  $\omega$ determining the orientation of the light-front plane, only if this orientation is varied relative to orientation of the particle momenta. 
If all the momenta are zero, the dependence on orientation of the LF plane, that is, on $\omega$ also disappears. For example, the variables 
$\vec{R}_{\perp},x$, Eq. (\ref{Rx}), are defined relative to the LF orientation. If $\vec{k}=0$, then the values $\vec{R}_{\perp}=0,\,x=1/2$ don’t depend on the LF orientation. We also expect that not only the scalar Fock components cease to depend on $\omega$ but also the spin structures containing the terms like
$\sim Y_{lm}(\vec{n})$ (or similar structures in the case of constituents with non-zero spin) should not appear. 
 Namely this latter condition will be applied below to the two-body wave function.  
In this way, we will find the coefficients $c_a$ in the superpositions $\Phi=\sum_a c_a\Phi_a$ corresponding to $l=1$ and $l=2$. Accuracy of the result will be checked by comparison with solutions extracted from corresponding BS amplitudes.

 %%%%%%%%%%%%
\subsection{$l=1$}  \label{psi_1}
In this case, the physical solution is the superposition of two solutions Eq. (\ref{bspsi1}) with $a=0$ and $a=1$:
\begin{equation}\label{psiph1}
 \psi_{1m}(\vec{k},\vec{n})=c_0 g_1^{(0)}(k,\theta)\chi_{1m}^{(0)}(\vec{k},\vec{n})
+c_1 g_1^{(1)}(k,\theta)\chi_{1m}^{(1)}(\vec{k},\vec{n}).
\end{equation}
As explained above, we impose the following natural physical requirement: if the relative momentum of particles is taken zero, the dependence of LFWF on $\vec{n}$ should also disappear since in this case its orientation is not related to any physical direction and cannot make any influence on the wave function.  
This means that LFWF $\psi_{1m}(\vec{k}\to 0,\vec{n})$ should not depend on $\vec{n}$. That is, the terms $\propto Y_{1m}(\vec{n})$ in (\ref{psiph1}) should cancel each other.
It happens if  $c_0 g_1^{(0)}(k,\cos\theta)=c_1 g_1^{(1)}(k,\cos\theta)$ at $k\to 0$. Hence,
\begin{equation}
    \label{C0C1}
    c_0=c_1\lim_{k\to 0}\frac{g_1^{(1)}(k,\cos\theta)}{g_1^{(0)}(k,\cos\theta)}.
\end{equation}
This requirement together with the normalization condition determines the coefficients $c_0,c_1$.

%%%%%%%%%%%%%
\subsection{$l=2$}\label{psi_2}
Now the physical solution is the superposition of three solutions Eq. (\ref{psi2a}) with $a=0,1$ and $a=2$:
\begin{eqnarray}\label{psiph2}
 \psi_{2m}(\vec{k},\vec{n})&=&c_0 g_2^{(0)}(k,\cos\theta)C_0\chi_{2m}^{(0)}(\vec{k},\vec{n})
 \nonumber\\
&+&c_1 g_2^{(1)}(k,\cos\theta)C_1\chi_{2m}^{(1)}(\vec{k},\vec{n})
\nonumber\\
&+&c_2 g_2^{(1)}(k,\cos\theta)C_2\chi_{2m}^{(1)}(\vec{k},\vec{n}).
\end{eqnarray}
The common eigenfunctions $\chi_{2m}^{(a)}(\vec{k},\vec{n})$ of the operators $\vec{L}^2,L_z,A^2$ for $l=2$  are defined in Eq. (\ref{chiL2}). 
Inserting their explicit term into (\ref{psiph2}), we find the  $\psi_{2m}(\vec{k}\to 0,\vec{n})$ becomes $\vec{n}$-independent if
\begin{equation}\label{c01}
c_0 =c_2\lim_{k\to 0}\frac{g_2^{(2)}(k,\cos\theta)}{g_2^{(0)}(k,\cos\theta)},\quad
c_1 =c_2 \lim_{k\to 0}\frac{g_2^{(2)}(k,\cos\theta)}{g_2^{(1)}(k,\cos\theta)}.
\end{equation}
Note that introduction of the coefficients $C_{0,1,2}$ into (\ref{psiph2}) simplifies the conditions (\ref{c01}).

\section{LFWF via BS amplitude}
\begin{figure}[!ht]
\centering
\includegraphics{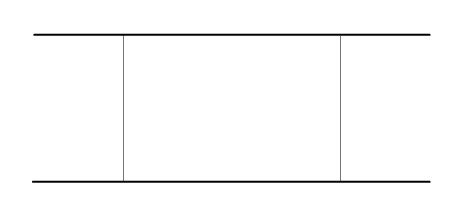}
\caption[*]{Feynman ladder graph with two exchanges.}
\label{feyn2}
\end{figure}
To check our proposal for the physical LFWF based on the superposition of the states with different $a$, we will find it by another, more lengthy and difficult but undoubted way.
Namely, we will solve the BS equation with the one-boson exchange kernel and project its solution on the LF plane. As well known, this projection provides the LFWF (see e.g. \cite{cdkm}, Sec. 3.3). 
Projection of the BS amplitude on the LF plane means that in the BS amplitude $\Phi_{BS}(x_1,x_2,p)$ in the coordinate space ($x_1$, $x_2$ are the four-vectors) we constrain the arguments $x_1, \,x_2$ by the LF plain $t+z=0$ and then go to the momentum space (make the Fourier transform). The result is reduced to the integral over $k_-$ from the BS amplitude $\Phi_{BS}(k,p)$. In the explicitly covariant version these constraints are achieved by multiplying the BS amplitude by the delta-functions $\delta(\omega\cd x_1)\delta(\omega\cd x_2)$. 
Making then the Fourier transform, we find that 
LFWF  is obtained by application to the BS amplitude $\Phi_{BS}(k,p)$ the following operation \cite{bs1}: 
\begin{equation}
\label{betatrans}
    \begin{split}
      \psi=\frac{x(1-x)}{\pi}\int_{-\infty}^{\infty}\Phi_{BS}\left(k+\beta \frac{\omega}{(p\cdot \omega)},p\right)d\beta.
    \end{split}
\end{equation}

This method does not suffer from violation of the rotational invariance. As explained above, this violation is a consequence of truncation. We mean, however, not any arbitrary truncation. Given Feynman graph is represented as sum of the LF-time ordered graphs.  The Feynman graphs which
topologically are not equivalent to each other generate different groups of the time ordered graphs. The members from different groups cannot be reduced to each other by the time ordering. The truncation not violating the rotational invariance
means omitting these topologically different groups but keeping all the graphs belonging to one group, member of which are obtained by different time orderings of a given Feynman graph.
Truncation omitting part of the time ordered graphs from a given group violate rotational invariance.
\begin{figure}[!ht]
\centering
 \includegraphics{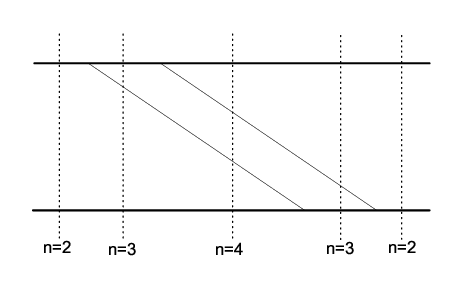}
\caption[*]{One of the six time-ordered ladder graphs, generated by
the Feynman graph Fig.~\protect{\ref{feyn2}}.}
 \label{LF2}
\end{figure}

Solution of the BS equation with the one boson exchange  kernel automatically takes into account infinite sum of these exchanges corresponding to different orders of the ladder graphs. For example,  the second order ladder Feynman graph is shown in Fig. \ref{feyn2}. In terms of the LF time-ordered graphs, the Feynman graph Fig. \ref{feyn2} corresponds to six  time-ordered graphs. One of them is shown in Fig. \ref{LF2}.
Just the graphs like Fig. \ref{LF2} are omitted in the LF one boson exchange kernel since, from point of view of the time-ordered graphs, they contain two bosons in the intermediate state.  Their neglect leads to violation of the rotational invariance.  These graphs (and with more bosons in the intermediate state, up to infinity) are effectively incorporated in the Feynman one boson exchange kernel that preserves the rotational invariance.
Comparison of the superposition of LFWF’s found from  the LF equation with projection of the BS amplitude on the LF plane will allow us to estimate the accuracy of the LFWF’s found directly from the LF equation.

To deal with the BS amplitude, we represent it via Nakanishi integral \cite{Nakanishi1963, Nakanishi1969}.
For the non-zero angular momentum $l$, the BS amplitude in the rest frame $\vec{p}=0$ is proportional to the spherical function $Y_{lm}(\hat{\vec{k}})$:
\begin{widetext}
\begin{equation}
\label{PhiL}
      \Phi_{BS}(k,p)=i|\Vec{k}|^l Y_{lm}(\hat{\vec{k}})\int_{-1}^{1}d\xi’\int_0^{\infty}d\gamma'\frac{g(\gamma’,\xi’)}
      {[k^2+p\cd k\, \xi’-\gamma'-m^2+M^2/4+i\epsilon]^n},
\end{equation}
\end{widetext}
where $n=3+l$. 
Vector $\vec{k}$ is the spatial part of the four-vector \mbox{$k=\frac{1}{2}(k_1-k_2)$} in the rest system $\vec{p}=0$.
The sign ”hat” (ˆ) over $\vec{k}$ means that $\hat{\vec{k}}$ is the unit vector.   
 In general frame,
the argument $\vec{k}$ is 
\begin{equation}
\label{ktrans}
      \tilde{\vec{k}} =L^{-1}(p)\vec{k}= \vec{k}-\frac{\vec{p}}{M}\left[k_0-\frac{(\vec{k},\vec{p})}{M+p_0}\right],
\end{equation}
where $L^{-1}(p)$ is the Lorentz transformation to the rest frame with $\vec{p}=0$, providing $L^{-1}(p)\vec{p}=0$.
BS equation reads:
\begin{eqnarray}\label{bs}
\Phi_{BS}(k,p)&=&\frac{i^2}{\left[(\frac{p}{2}+k)^2-m^2+i\epsilon\right]
\left[(\frac{p}{2}-k)^2-m^2+i\epsilon\right]}
\nonumber\\
&\times&\int \frac{d^4k'}{(2\pi)^4}iK(k,k',p)\Phi_{BS}(k’,p).
\end{eqnarray}
Following to Refs. \cite{bs1,nak2},
 we will derive from Eqs.  (\ref{bs}) and (\ref{PhiL}) the equation for the Nakanishi function $g(\gamma,\xi)$. This derivation consists of two principal steps. 
 
 As the first step, we substitute into Eq. (\ref{bs}) the BS amplitude in the form (\ref{PhiL}) and apply to both parts of equation  the operation (\ref{betatrans}), i.e., take
the LF projection, like it was done in Ref. \cite{bs1}.
This means that we replace the argument as 
$$
k\to k+\beta \frac{\omega}{(p\cdot \omega)}
$$
and integrate over $\beta$. In this way, we get the equation:
\begin{eqnarray} \label{bsnew}
&&\int_0^{\infty}\frac{g(\gamma',\xi)d\gamma'}{\Bigl[\gamma'+\gamma +\xi^2 m^2+(1-\xi^2)\kappa^2\Bigr]^{n-1}}=G_n(\gamma,\xi)\equiv
\nonumber\\
&&\int_0^{\infty}d\gamma'\int_{-1}^{1}d\xi’\;V_n(\gamma,\xi;\gamma',\xi') g(\gamma',\xi’),
\end{eqnarray}
where  
\begin{widetext}
\begin{eqnarray}\label{eq2}
G_n(\gamma,\xi)&=&
\int_{-1}^\xi d\xi’
\int_0^{\infty}d\gamma’\frac{\alpha m^2 g(\gamma’,\xi’)(1-\xi)^{n-1}}{2\pi(\gamma+\xi^2m^2+\kappa^2(1-\xi^2))}
\left[\int_0^1\frac{(1-v)^lv^{n-1}dv}{D^{n-1}}\right]
\nonumber\\
&+&
\int_{\xi}^1 d\xi’
\int_0^{\infty}d\gamma’\frac{\alpha m^2 g(\gamma’,\xi’)(1+\xi)^{n-1}}{2\pi(\gamma+\xi^2m^2+\kappa^2(1-\xi^2))}
\left[\int_\xi^1\frac{(1-v)^lv^{n-1}dv}{D^{n-1}(\xi\to-\xi,\xi’\to -\xi’)}\right]
\end{eqnarray}
with
\begin{eqnarray}\label{D}
D&=&v(1-v)(1-\xi’)\gamma + v(1-\xi)\gamma’
+v(1-\xi)(1-\xi’)[1+\xi(1-v) + v\xi’]\kappa^2
\nonumber\\
&+&v[(1-v)(1-\xi’)\xi^2 + v{\xi’}^2(1-\xi)]m^2
+(1-v)(1-\xi)\mu^2.
\end{eqnarray}
\end{widetext}

For the S-wave ($l=0,\,n=3$) the kernel $V_3$ in given explicitly in Ref. \cite{bs1}.

The equation (\ref{bsnew}) is solvable numerically, for $l=0$ it was solved in \cite{bs1}. However, since it contains the integrals in both parts, finding stable solutions sometimes requires extra efforts. In \cite{nak2} it was noticed that the integral in the left-hand side of Eq. (\ref{bsnew}) is the Stieltjes transform which, according to \cite{Sumner_49,schwarz}, can be inverted. Namely, solution of Eq. (\ref{bsnew}) relative to $g(\gamma,\xi)$ can be represented in the form:
\begin{eqnarray}\label{ggz}
g(\gamma,\xi)&=&\frac{(n-2)\gamma^{n-2}}{2\pi }\int_{-\pi+\epsilon}^{\pi-\epsilon}(1+e^{i\phi})^{n-3}
\\
&\times&G_n(\gamma e^{i\phi}-\xi^2m^2-\kappa^2(1-\xi^2),\xi) e^{i\phi}d\phi,
\nonumber
\end{eqnarray}
where $\epsilon\to 0$. With $G_n(\gamma,\xi)$ defined in (\ref{eq2}), the integral (\ref{ggz}) over $\phi$ can be calculated analytically.
Since, according to Eq. (\ref{bsnew}), $G_n(\gamma,\xi)$ is integral containing $g(\gamma,\xi)$, 
substituting in (\ref{ggz}) instead of $G_n$ the r.h.-side of Eq. (\ref{bsnew}),  we reduce it to the standard integral equation
for the Nakanishi function:
\begin{equation}\label{gNg}
g(\gamma,\xi)=\int_0^{\infty}d\gamma’\int_{-1}^1d\xi’ N(\gamma,\xi,\gamma’,\xi’)g(\gamma’,\xi’).
\end{equation}
For $l=0$, expression for the kernel $N$ was derived in \cite{nak2} and, in different but mathematically equivalent form, in \cite{VK-Nak}. This allowed to make finding the Nakanishi function $g(\gamma,\xi)$ and, via it, LFWF more stable and reliable. 

The second step is, with given $g(\gamma,\xi)$, to find the LFWF by Eq. (\ref{betatrans}). Substituting in (\ref{betatrans}) the BS amplitude in the form (\ref{PhiL}) for $l=0$, we find LFWF:
\begin{equation}\label{psi-g}
\psi(k_\perp,x)=\frac{1}{\sqrt{4\pi}}\int_0^{\infty}\frac{x(1-x)g(\gamma',1-2x)d\gamma'}
{\Bigl[\gamma'+k_\perp^2 +m^2-x(1-x)M^2\Bigr]^2}.
\end{equation}
For calculating the integral (\ref{betatrans}) for $l=0$ we used the formula:
\begin{equation}\label{int0}
\int_{-\infty}^{\infty}\frac{d\beta}{(\beta x-y-i\epsilon)^3}=\frac{i\pi}{y^2}\delta(x).
\end{equation}
Similarly, substituting in (\ref{betatrans}) the BS amplitude the function $\Phi_{BS}(k, p)$ for $l=1$ and $l=2$, we find corresponding  LFWF’s, presented in Appendix \ref{app2}.
For $l=1$ the numerator in (\ref{betatrans}) contains extra degree of $\beta$ and the degree of denominator is 4.  In this case, to calculate the integral, we still need the formula (\ref{int0}) as well as similar formula with extra $\beta$ in the numerator and extra degree of denominator. The latter one is obtained from (\ref{int0}) by taking the derivative over $x$ from both parts. It contains the derivative $\delta’(x)$. In its turn, this results in the derivative $g’_{\tilde{z}}(\gamma’,\tilde{z})$ in LFWF, Eq. (\ref{psiBSL1a}). Similarly, the second derivative $g^{\prime\prime}_{\tilde{z}\tilde{z}}(\gamma’,\tilde{z})$ appears in Eq. (\ref{psiBSL2a}). 

Note that the functions (\ref{psiBSL1a}) (for $l=1$) and (\ref{psiBSL2a}) (for $l=2$) are automatically obtained in the form of superpositions 
of the functions $\chi_{1m}^{(0,1)}$ and  $\chi_{2m}^{(0,1,2)}$ respectively.

\section{Numerical results}\label{num}

\subsection{$l=1$}

We will perform a numerical comparison between  the functions $\psi_{1M}$ \eqref{psiph1} and $\psi^{BS}_{1M}$ \eqref{psiBSL1a} and the corresponding bound state masses. Our goal is to verify whether these functions exhibit the same behaviour for fixed values of $m$.

\begin{figure}[h!]
\begin{center}
\includegraphics[width=0.4\textwidth]{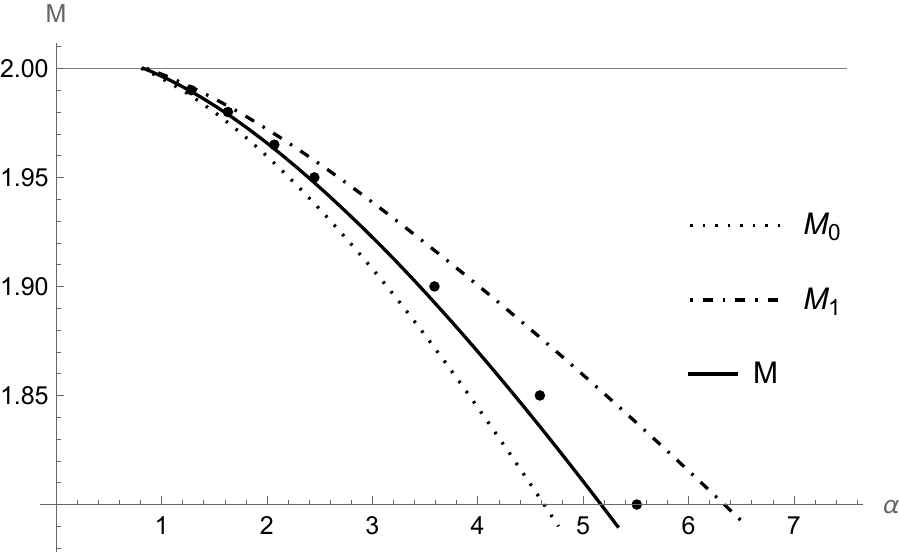}
\end{center}
\caption{$l=1$. Masses as functions of $\alpha$ for $\mu=0.07$. BS - dots, $M_0$ - dotted line, $M_1$ - dot-dashed line, $M$ - solid line.}\label{ML1}
\end{figure}

The state described by the wave function which is  the sum \eqref{psiph1} of two orthogonal terms corresponds to the mass squared
\begin{equation}\label{Mc1}
M^2=c_0^2M_0^2+c_1^2M_1^2,
\end{equation}
where $M_{0,1}$ are the masses determined by the equations \eqref{eq10} and \eqref{eq11}, and normalization condition is $c_0^2+c_1^2=1$.  The mass of this state as function of the coupling constant $\alpha$ is shown in Fig. \ref{ML1}.
This mass is rather close to the value obtained from the Bethe-Salpeter equation.

For comparison of the wave functions, we take the projection  $m=0$ and direct $\vec{k}\parallel z$. In this configuration, the functions $\chi_{10}^{(a)}$, Eq. \eqref{chiL1}, obtain the form:
\begin{eqnarray}
\label{chi10_01}
\chi^{(0)}_{10}(\vec{k},\vec{n})&=&\frac{1}{2}\sqrt{\frac{3}{\pi}}z^2,
\nonumber\\
\chi^{(1)}_{10}(\vec{k},\vec{n})&=&\frac{1}{2}\sqrt{\frac{3}{\pi}}(1-z^2).
\end{eqnarray}

\begin{figure}[h!]
\begin{center}
\includegraphics[width=0.4\textwidth]{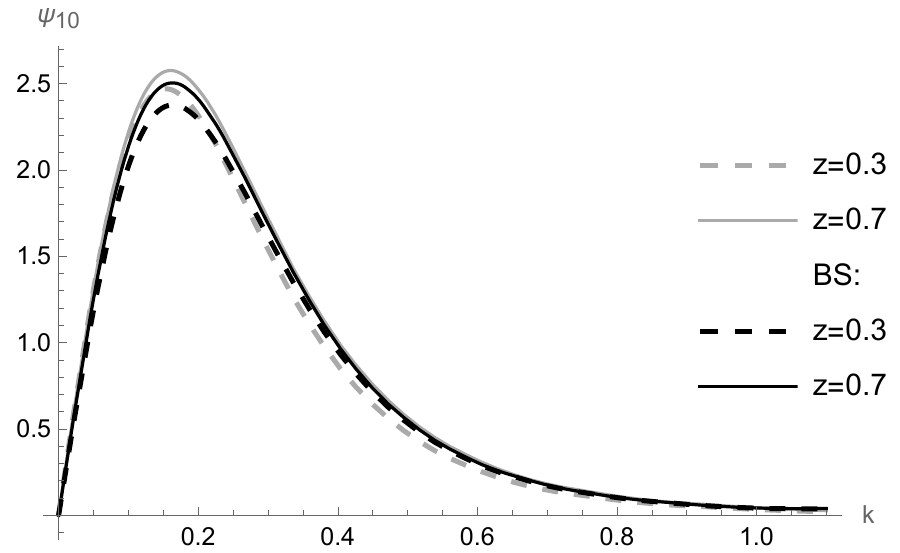}
\includegraphics[width=0.4\textwidth]{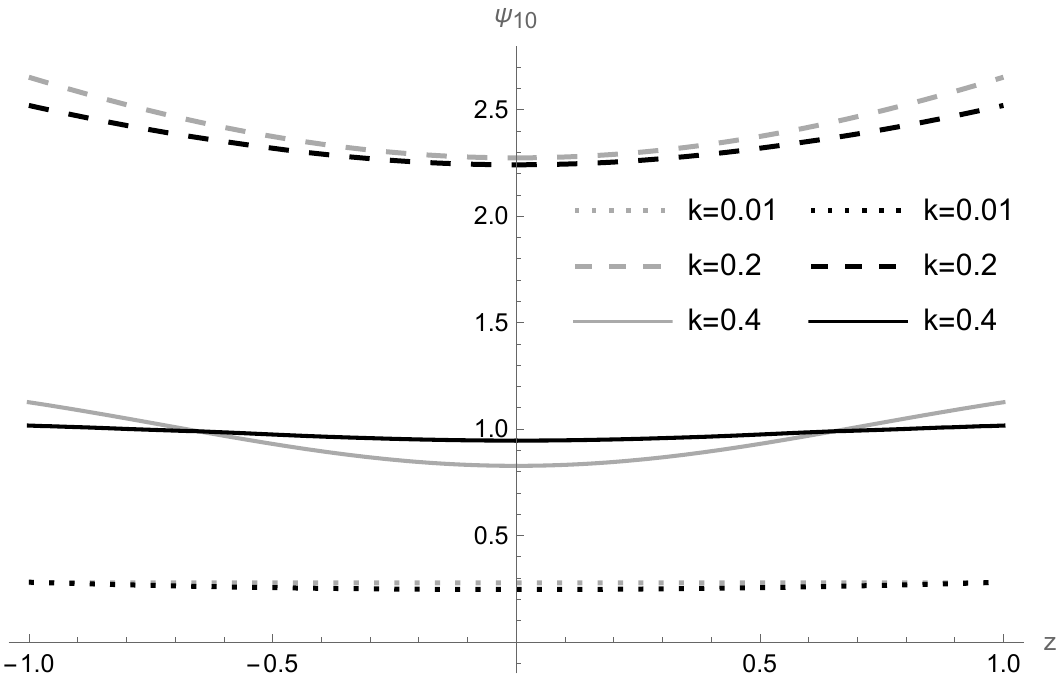}
\end{center}
\caption{$B_{BS}=0.1$, $\alpha=3.59$, $m=0$, $c_0$ and $c_1$ from the formula \eqref{C0C1}. $\psi_{10}$ from \eqref{psiph1}  - gray lines, 
$\psi^{BS}_{10}$ from  \eqref{psiBSL1a} - black lines. \\
Top: functions of $k$. $z=0.3,0.7$.\\
Bottom: functions of $z$. $k=0.01,0.2,0.4$. }\label{CompPsiL10}
\end{figure}

The plots of the functions $\psi_{10}$ \eqref{psiph1} and $\psi^{BS}_{10}$ \eqref{psiBSL1a} for $\alpha=3.59$ and $\mu=0.07$ are shown in Fig. \ref{CompPsiL10}. This value of the coupling constant $\alpha$ corresponds to the binding energy for Bethe–Salpeter equation $B_{BS}=0.1$, and $B_0=0.128$,  $B_1=0.083$ for equations \eqref{eq10} and \eqref{eq11}.

We find $c_0$ and $c_1$ from \eqref{C0C1}.
The normalization conditions $\int |\psi_{10}(k,z)|^2d^3\vec{k}=\int |\psi^{BS}_{10}(k,z)|^2d^3\vec{k}=1$ and $\int \sum_{m}|g_1^{(a)}(k,\theta)|^2\chi_{1m}^{(a)*}(\hat{\vec{k}})\chi_{1m}^{(a)}(\hat{\vec{k}}) d^3\vec{k}=1$ are used on the plots.

\begin{figure}[h!]
\begin{center}
\includegraphics[width=0.4\textwidth]{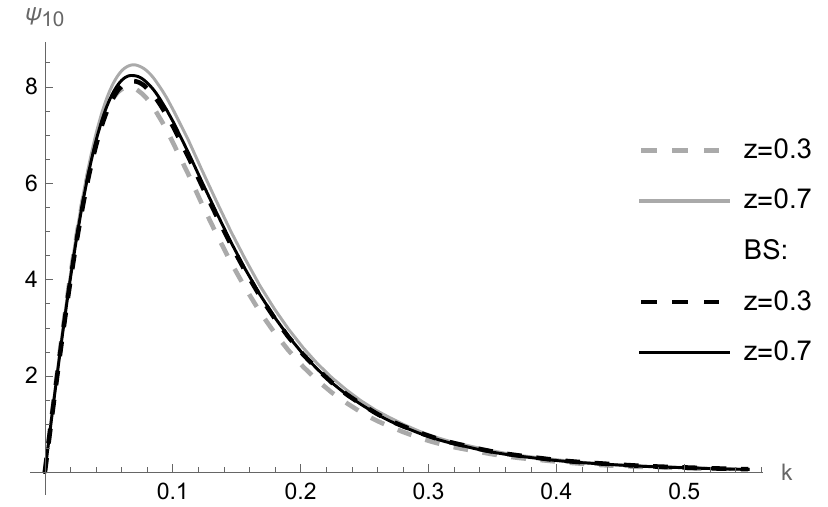}
\end{center}
\caption{$B_{BS}=0.01$, $\alpha=1.28$, $m=0$, $c_0$ and $c_1$ from analytical formula \eqref{C0C1}. $\psi_{10}$  from \eqref{psiph1}    - gray lines, $\psi^{BS}_{10}$ from  \eqref{psiBSL1a} - black lines. 
 Functions of $k$. $z=0.3,0.7$.}
\label{CompPsiL10sm}
\end{figure}

For the lower value of coupling constant we compare \eqref{psiph1} with \eqref{psiBSL1a} on Fig. \ref{CompPsiL10sm}  by plotting them as functions of $k$, $z$. 
The coupling constant $\alpha=1.28$ corresponds to $B_{BS}=0.01$ and to the binding energies $B_0=0.0126$ and $B_1=0.0086$ for the equations \eqref{eq10} and \eqref{eq11} and the same $\mu=0.07$. We take $c_0$ and $c_1$ from the formula \eqref{C0C1}.

It can be noticed that a reduction in the value of $\alpha$ leads to some improvement in the agreement between the curves in Fig. \ref{CompPsiL10sm}.

\subsection{$l=2$}

In this subsection, analogously to the previous one, we aim to carry out a numerical comparison of the functions $\psi_{2m}$, Eq. \eqref{psiph2}, with 
$\psi^{BS}_{2m}$, Eq. \eqref{psiBSL2a}, and the corresponding bound state masses.

\begin{figure}[h!]
\begin{center}
\includegraphics[width=0.4\textwidth]{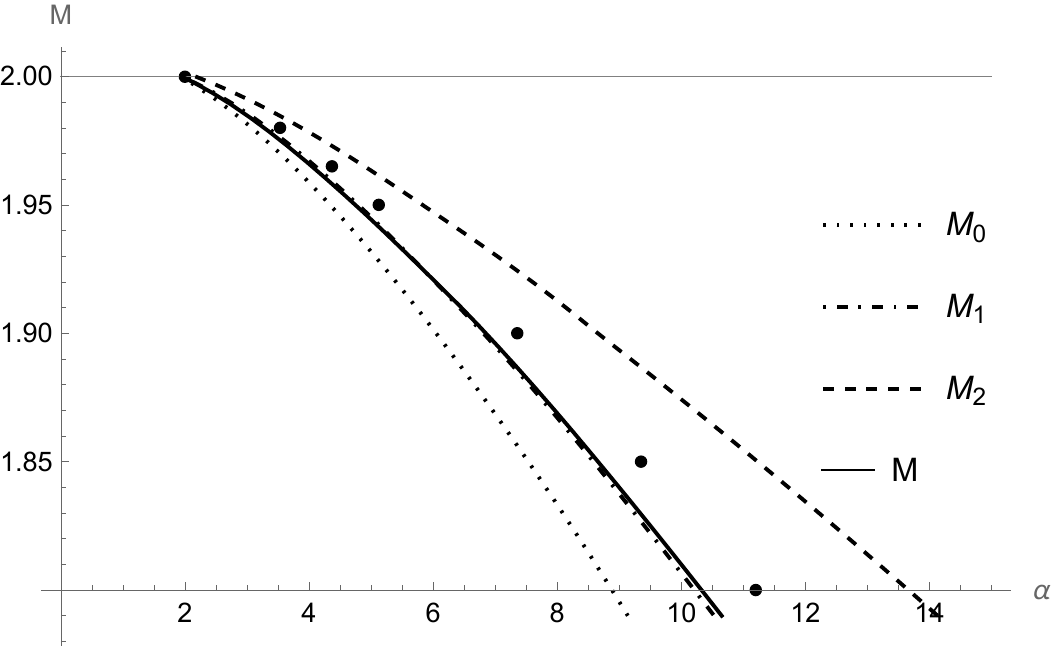}
\end{center}
\caption{$l=2$. Masses as functions of $\alpha$ for $\mu=0.07$. BS - dots, $M_0$ - dotted line, $M_1$ - dot-dashed line, 
$M_2$ - dashed line, $M$ - solid line.}\label{ML2}
\end{figure}

Similarly to the case $l=1$, the state described by the wave function which is  the sum \eqref{psiph2} of three orthogonal terms corresponds to the following mass squared
\begin{equation}\label{Mc2}
M^2=c_0^2M_0^2+c_1^2M_1^2+c_2^2M_2^2,
\end{equation}
where $M_{0,1,2}$ are the masses determined by the equations (\ref{eqL20c}-  \ref{eqL22c}) and sum the coefficients squared is normalized $c_0^2+c_1^2+c_2^2=1$. The mass of this state as functions of the coupling constant $\alpha$ is shown in Fig. \ref{ML2}.
This mass is rather close to the value obtained from the Bethe-Salpeter equation for small values of $\alpha$.

We take the function $\psi_{2m}(\vec{k},\vec{n})$ from \eqref{psiph2} with $m=0$, direct $\vec{k}\parallel z$, and compare it with \eqref{psiBSL2a}. The functions $\chi_{20}^{(a)}$ \eqref{chiL2} obtain the form:
\begin{eqnarray}
\label{}
\chi^{(0)}_{20}(\vec{k},\vec{n})&=&\frac{\sqrt{5}}{2}(3{z}^2-1),
\nonumber\\
\chi^{(1)}_{20}(\vec{k},\vec{n})&=&3\sqrt{\frac{3}{2}}{z}(1-{z}^2),
\nonumber\\
\chi^{(2)}_{20}(\vec{k},\vec{n})&=&\frac{3}{4}\sqrt{5}(1-{z}^2)^2.
\nonumber
\end{eqnarray}

\begin{figure}[h!]
\begin{center}
\includegraphics[width=0.4\textwidth]{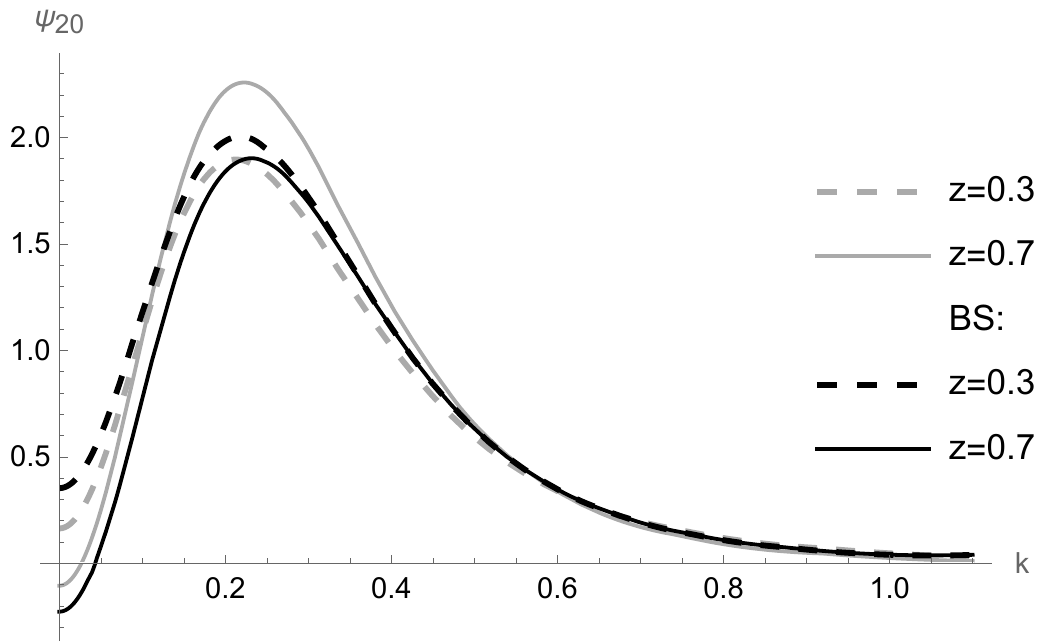}
\includegraphics[width=0.4\textwidth]{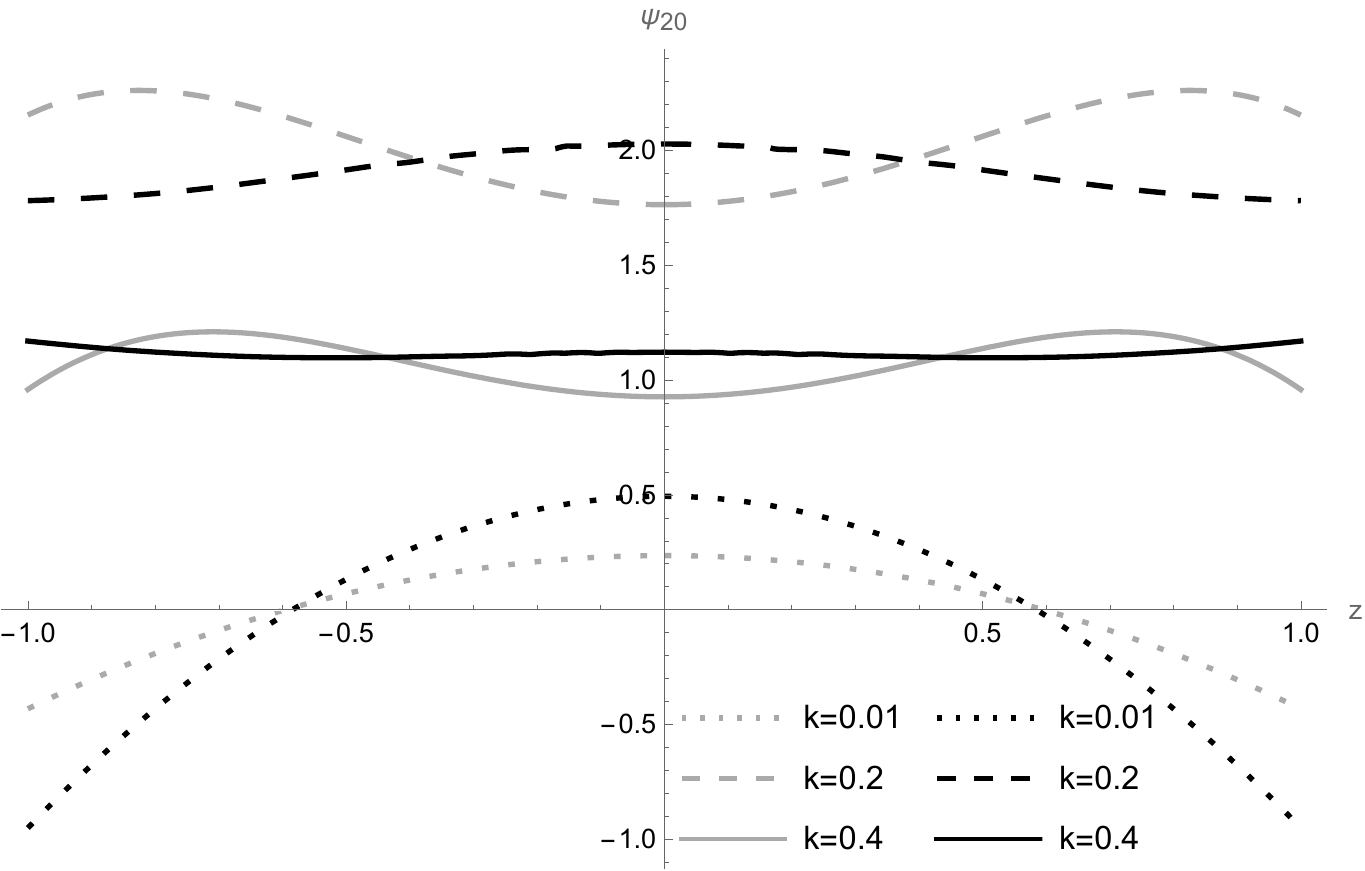}
\end{center}
\caption{$B_{BS}=0.1$, $\alpha=7.36$, $m=0$, $c_0$, $c_1$ and $c_2$ from analytical formula \eqref{c01}. $\psi_{20}$ from \eqref{psiph2}  - gray lines, $\psi^{BS}_{20}$ from  \eqref{psiBSL2a} - black lines. \\
Top: functions of $k$. $z=0.3, 0.7$.\\
Bottom: functions of $z$. $k=0.01, 0.2, 0.4$. }\label{CompPsiL20}
\end{figure}

The plots of the functions $\psi_{20}$ and $\psi^{BS}_{20}$ vs. $k$, $z$ for $\alpha=7.36$ and $\mu=0.07$ are shown in Fig. \ref{CompPsiL20}.  This value of the coupling constant $\alpha$ corresponds to the binding energy in the Bethe–Salpeter equation $B_{BS}=0.1$, and $B_0=0.144$, $B_1=0.115$ and $B_2=0.077$ for equations \eqref{eqL20c}-\eqref{eqL22c}. The numerical values of $c_2$, $c_1$ and $c_0$ follow from the formula \eqref{c01}.
    The normalization conditions  $\int |\psi_{2m}(k,{z})|^2d^3\vec{k}=\int |\psi^{BS}_{2m}(k,{z})|^2d^3\vec{k}=1$ and  $\int \sum_{m}|g_2^{(a)}(k,\theta)|^2\chi_{2m}^{(a)*}(\hat{\vec{k}})\chi_{2m}^{(a)}(\hat{\vec{k}}) d^3\vec{k}=1$ are used on the plots.

As can be observed from the Fig. \ref{CompPsiL20} the behavior of the functions $\psi_{2m}$ and $\psi^{BS}_{2m}$ in the case of $l = 2$ differs more significantly than in the case of $l = 1$. This might be related with less accuracy of numerical calculations in the case $l=2$.

\section{Conclusion}\label{concl}

Making, in the LFD framework, truncation and solving equation  for  LFWF (or system of equations), we get a spectrum of masses. They are different for different  projections of the angular momentum of the $z$-axis. This is a manifestation of violation of the rotational invariance. In addition, the question arises:
which solution from this spectrum should be used for description of the physical system? In the version of LFD with the LF plane defined by the equation 
$\omega\cd x=0$ the energies don’t depend anymore on projections of the angular momentum of the $z$-axis, but they
depend on projection of the angular momentum on the direction $\vec{\omega}$ determining orientation of the LF plane.

If the algebra of the Poincar\'e generators is satisfied (including the rotation subgroup), 
the rotational symmetry in LFD is ensured. This symmetry is violated if the Poincar\'e algebra is violated. When we truncate the Fock space -- make it finite, we also make the matrices representing the generators also finite and violate in this way the Poincar\'e algebra. In its turn, this results in violation of the rotational symmetry.

Without truncation, the masses $M_a$ corresponding to solutions with different projections $a$ of the angular momentum on $\vec{\omega}$  correspond to one and the same mass (they are degenerate), whereas the physical solution is the superposition of these solutions. The recipe proposed in \cite{NP2000,NP2001,NP2002} and confirmed on the present work is to keep this superposition  even after truncation. It becomes approximate (since making truncation, we make approximation). We have checked  above that this approximation works rather well. 

Therefore, in practice, we should solve the problem “as it is”, that is, to find the $a$-dependent spectrum and corresponding wave functions. Then, in spite of the fact that these solutions correspond to different energies, we should take the superposition of these solutions. The coefficients in this superposition are found from the condition that at zero momentum the wave function does not depend on orientation of the LF plane, that is, on the vector $\vec{n}$. Since the coefficients in this superposition are constants, they are determined from this condition unambiguously.

As it was shown, the method using as the basis found in Sec. \ref{eigenf}
 common eigenfunctions $\chi^{(a)}_{lm}(\vec{k},\vec{n})$ of the operators $\vec{L}^2,L_z, A^2$ provides, of course, the same mass spectrum as obtained from Eq. (\ref{eqlm}), representing the wave function in the form (\ref{psi_m}). The eigenvalue spectrum is a property of equation and does not depend on the method used to find it. Using the basis $\chi^{(a)}_{lm}(\vec{k},\vec{n})$, we get double advantage. 
({\it i}) We split the solutions in the groups with given $l$, labeled within group by $a$. Namely, these solutions form superpositions with given $l$. ({\it ii})
Explicit and simple dependence of the basis functions $\chi^{(a)}_{lm}(\vec{k},\vec{n})$ on $\vec{n}$ simplifies finding the superposition coefficients ensuring independence of the sum on $\vec{n}$ at zero relative momenta of the constituents (see examples in the subsections \ref{psi_1} and \ref{psi_2} in Sec. \ref{ph-sol}).

Comparison of these solutions with LFWF’s found by projection of the BS amplitudes on the LF plane shows good coincidence of the binding energies and of the wave functions themselves, that confirms the validity of the method. Though, deviation for $l=2$ is larger than for $l=1$. This may be due to a deterioration in numerical accuracy with increase of $l$. Note also that the one boson exchange kernels used in the LF equation and in the BS one are not completely identical. In the LF equation the kernel is exhausted by two diagrams Fig. \ref{obe_lf}. Whereas, the BS solution implicitly takes into account
all the time ordered graphs generated by the time ordering of all the ladder contributions, like, for example, shown in Fig. \ref{LF2}.  They are irreducible  and of higher order in the coupling constant.
Note that the dimensionless coupling constant is rather large: $\alpha=4\div 7$ that is quite beyond the perturbative domain. Therefore, the above mentioned good coincidence may come as a surprise. On the other hand, let us notice that the higher order graphs contain more particles in the intermediate states (in higher Fock sectors). This increase of the Fock sector content results in decrease of its contribution to the amplitude. The latter can give more strong effect than
increase of the coupling constant \cite{LKMV_PLB,klsv2016}.

LFWF’s, found in this way, with the approximately restored rotational symmetry, can be used for description of properties of hadrons and of high momentum components of nuclei, in particular, for calculating their electromagnetic form factors.

\appendix

\section{LFWF’s for the angular momenta $l=1$ and $l=2$ extracted from the BS amplitude}\label{app2}
Substituting in Eq. (\ref{betatrans}) the BS amplitude (\ref{PhiL}), taking $l=1,2$ and calculating the integral over $\beta$, we find the following LFWF’s
\begin{widetext}
$l=1$:
\begin{equation}
\label{psiBSL1a}
\begin{split}
       \psi^{BS}_{1m}=&\chi_{1m}^{(1)}\frac{1}{4}\left(1-\frac{k^2z^2}{k^2+m^2}\right)k\int_0^{\infty}d\gamma'{g(\gamma',\tilde{z})\over [\gamma'+\gamma+m^2\tilde{z}^2+\kappa^2(1-\tilde{z}^2)]^{n-1}}\frac{(-1)^{n+1}2}{(n-1)}+\\
       &+\chi_{1m}^{(0)}\frac{1}{4}\left(1-\frac{k^2z^2}{k^2+m^2}\right)\left[\frac{k}{\sqrt{k^2+m^2}}\frac{M}{2}\int_0^{\infty}d\gamma'{g(\gamma',\tilde{z})\over [\gamma'+\gamma+m^2\tilde{z}^2+\kappa^2(1-\tilde{z}^2)]^{n-1}}\frac{(-1)^{n+1}2}{(n-1)}-\right.\\
       &\left.-\frac{1}{Mz}\int_0^{\infty}d\gamma’{g’_{\tilde{z}}(\gamma',\tilde{z})\over [\gamma'+\gamma+m^2\tilde{z}^2+\kappa^2(1-\tilde{z}^2)]^{n-2}}\frac{(-1)^{n+1}2}{(n-2)(n-1)}\right], 
    \end{split}
\end{equation}

where $\gamma=\vec{k}_{\perp}^2=\Vec{k}^2-(\Vec{n}\cd\Vec{k})^2=k^2(1-z^2)$, $\tilde{z}=1-2x=(\Vec{n}\cd\Vec{k})/\sqrt{\Vec{k}^2+m^2}=(zk)/\sqrt{\Vec{k}^2+m^2}$, 
$z=(\Vec{n}\cd\Vec{k})/k$, $s=4\epsilon_k^2=4(\Vec{k}^2+m^2)$. To avoid confusion, we emphasize that it is LFWF, whereas the subscript “$^{BS}$” means that it was extracted from the BS amplitude.

For $l=2$:
    \begin{equation}
\label{psiBSL2a}
\begin{split}
\psi{^{BS}_{2m}}\sim&\chi_{2m}^{(2)}\frac{1}{\sqrt{4\pi}}\left(1-\frac{k^2z^2}{k^2+m^2}\right)k^2\int_0^{\infty}d\gamma'{g(\gamma',\tilde{z})\over [\gamma'+\gamma+m^2\tilde{z}^2+\kappa^2(1-\tilde{z}^2)]^{n-1}}\frac{(-1)^{n+1}2}{(n-1)}+\\
       &+\chi_{2m}^{(1)}\frac{1}{\sqrt{4\pi}}\sqrt{\frac{10}{3}}\left(1-\frac{k^2z^2}{k^2+m^2}\right)\frac{1}{z}\left[\frac{k^2}{\sqrt{k^2+m^2}}\frac{{\mathcal{M}}}{2}\int_0^{\infty}d\gamma'{g(\gamma',\tilde{z})\over [\gamma'+\gamma+m^2\tilde{z}^2+\kappa^2(1-\tilde{z}^2)]^{n-1}}\frac{(-1)^{n+1}2}{(n-1)}-\right.\\
       &\left.-\frac{k}{{\mathcal{M}}z}\int_0^{\infty}d\gamma’{g’_{\tilde{z}}(\gamma',\tilde{z})\over [\gamma'+\gamma+m^2\tilde{z}^2+\kappa^2(1-\tilde{z}^2)]^{n-2}}\frac{(-1)^{n+1}2}{(n-2)(n-1)}\right]+\\
       &+\chi_{2m}^{(0)}\frac{1}{\sqrt{4\pi}}\left(1-\frac{k^2z^2}{k^2+m^2}\right)\left[k^2\left(\frac{{\mathcal{M}}^2z^2}{4(k^2+m^2)}-\frac{(1-z^2)}{2}\right)\int_0^{\infty}d\gamma'{g(\gamma',\tilde{z})\over [\gamma'+\gamma+m^2\tilde{z}^2+\kappa^2(1-\tilde{z}^2)]^{n-1}}\frac{(-1)^{n+1}2}{(n-1)}-\right.\\
    &-\frac{kz}{\sqrt{k^2+m^2}}\int_0^{\infty}d\gamma’{g’_{\tilde{z}}(\gamma',\tilde{z})\over [\gamma'+\gamma+m^2\tilde{z}^2+\kappa^2(1-\tilde{z}^2)]^{n-2}}\frac{(-1)^{n+1}2}{(n-2)(n-1)}+
    \\
    &\left.+{1\over {\mathcal{M}}^2}\int_0^{\infty}d\gamma'\frac{g^{\prime\prime}_{\tilde{z}\tilde{z}}(\gamma',\tilde{z})}{[\gamma'+\gamma+m^2{\tilde{z}}^2+\kappa^2(1-{\tilde{z}}^2)]^{n-3}}\frac{(-1)^{n+1}2}{(n-3)(n-2)(n-1)}\right]. 
    \end{split}
\end{equation}
\end{widetext}

% \begin{thebibliography}{10}
% \input{references.tex}
% \end{thebibliography}

\bibliographystyle{apsrev4-2}
\bibliography{references}
\end{document}